%% file: main.tex
\documentclass[journal]{IEEEtran}

\IEEEpubid{
\begin{minipage}{\textwidth}\centering
\vspace{7em}
\footnotesize
\textbf{Acknowledgement: This preprint has not undergone
peer review (when applicable) or any post-submission improvements or corrections. The Version of Record of this article is
published in SN Computer Science, and is available online at https://doi.org/10.1007/s42979-026-05087-1}
\end{minipage}
}

\usepackage{cite}
\usepackage{amsmath,amssymb,amsfonts}
\usepackage{algorithm, algpseudocode}
\usepackage{graphicx}
\usepackage{subfigure}
\usepackage{textcomp}
\usepackage{longtable}
\usepackage{xcolor}
\usepackage[utf8]{inputenc}
\usepackage{mathtools}
\usepackage[inkscapearea=page]{svg}
\usepackage{tikz}
\usetikzlibrary{shapes.geometric, arrows}
\def\BibTeX{{\rm B\kern-.05em{\sc i\kern-.025em b}\kern-.08em
    T\kern-.1667em\lower.7ex\hbox{E}\kern-.125emX}}

\begin{document}
\title{An Efficient Recommendation Filtering-based Trust Model for Securing Internet of Things}

\author{\IEEEauthorblockN{Muhammad Ibn Ziauddin, Rownak Rahad Rabbi, SM Mehrab, Fardin Faiyaz, Mosarrat Jahan}$^{*}$\\
\IEEEauthorblockA{Department of Computer Science and Engineering, University of Dhaka, Dhaka, Bangladesh\\
Email: muhammadibn-2017215011@cs.du.ac.bd, 2017-615-017@student.cse.du.ac.bd, smmehrabul-2017614964@cs.du.ac.bd, fardin-2020515627@cs.du.ac.bd,
mosarratjahan@cse.du.ac.bd}
\thanks{$^{*}$ Corresponding Author.}
}

\maketitle
\begin{abstract}
Trust computation is crucial for ensuring the security of the Internet of Things (IoT). However, current trust-based mechanisms for IoT have limitations that impact data security. Sliding window-based trust schemes cannot ensure reliable trust computation due to their inability to select appropriate window lengths. Besides, recent trust scores are emphasized when considering the effect of time on trust. This can cause a sudden change in overall trust score based on recent behavior, potentially misinterpreting an honest service provider as malicious and vice versa. Moreover, clustering mechanisms used to filter recommendations in trust computation often lead to slower results. In this paper, we propose a robust trust model to address these limitations. The proposed approach determines the window length dynamically to guarantee accurate trust computation. It uses the harmonic mean of average trust score and time to prevent sudden fluctuations in trust scores. Additionally, an efficient personalized subspace clustering algorithm is used to exclude recommendations. We present a security analysis demonstrating the resiliency of the proposed scheme against bad-mouthing, ballot-stuffing, and on-off attacks.  The proposed scheme demonstrates a competitive performance in detecting bad-mouthing attacks, while outperforming existing works with an approximately 44\% improvement in accuracy for detecting on-off attacks. It maintains its effectiveness even when the percentage of on-off attackers increases and in scenarios where multiple attacks occur simultaneously. Additionally, the proposed scheme reduces the recommendation filtering time by 95\%.
\end{abstract}


\input{introduction.tex}
\input{related_work}

\input{proposed_scheme}

\input{security_analysis}

\input{experimental_result}

\input{conclusion}

\bibliographystyle{IEEEtran}
\bibliography{IEEEabrv, main}
\end{document}

%% file: introduction.tex
\section{Introduction}

\IEEEPARstart{T}he Internet of Things (IoT) has brought greater comfort and ease in our daily lives by connecting the physical world to the Internet and enabling automated services. It facilitates the collection, communication, and aggregation of enormous amounts of data from massive low-powered sensing devices embedded in our surrounding environment. This vast IoT data is utilized by various service providers to provide better
amenities to users. With the reduction of human labor and human faults and the efficiency of automated services, IoT is rapidly becoming an integral part of our daily lives, connecting an ever-increasing number of devices ranging from wearables to smart homes, connected cars, healthcare, and many more to the internet \cite{iot_health}, \cite{iot_smarthome}, \cite{fixed_window_new_1}. However, despite numerous
advantages, the security and privacy concerns of IoT persistently prevent users from fully embracing this technology.

Due to the great potential of IoT, it is subject to different security attacks, leading to data breaches, financial losses, and other negative consequences \cite{iot_attackhome}, \cite{iot_attack_medical}. IoT security mechanisms are often inadequate due to limited processing power, storage capacity, energy constraints, and heterogeneity of devices. Besides, many IoT devices are designed with a focus on cost-effectiveness and ease of use, which often compromises security. Hence, it becomes essential to develop lightweight and efficient security mechanisms for IoT.

The notion of trust is a cost-effective security solution for IoT due to its low computational overhead \cite{trust_iot_reason1}, \cite{trust_iot_reason2}. Essentially, trust refers to a rating for a device to determine its trustworthiness, authenticity, importance, or recent behavior \cite{pki_vs_trust}. Trust-based solutions have proven effective in different domains such as vehicular ad hoc network (VANET) \cite{vnet_trust}, IoT \cite{servery_trust_1}, social IoT (SIoT) \cite{trust_in_SIOT}, collaborative IoT (CIoT) \cite{blockchain_trust_02}, and wireless sensor networks {(WSN)} \cite{fixed_sliding_window_01}. However, trust-based solutions for IoT have several limitations that impact the secure and reliable operation of the IoT environment, raising concern about data security. In sliding window-based trust management schemes, the window length determines the number of trust scores inside the window each time it slides over and controls the memorability of the previous behavior of an IoT device. An inappropriate window length hampers the reliable trust computation due to inadequate trust ratings inside the window, widening the attack surface for malicious entities. Besides, previous behavior history governed by window length helps to precisely analyze the behavior of an entity. Hence, window length is crucial for reliable trust computation. The situation becomes more critical when trust ratings inside the window are minimal due to the IoT device's infrequent operation, leading to difficulty in accurately analyzing the behavior of an entity. Existing sliding window-based trust management schemes that use fixed-length windows cannot guarantee sufficient trust ratings inside the trust window for reliable trust computation \cite{fixed_sliding_window_02}, \cite{fixed_sliding_window_01}, \cite{dynamic_window_wsn}. Although some research works address this issue using dynamic sliding window-based techniques  \cite{WSN_novel},\cite{ETERS},\cite{ETAS},\cite{dynamic_window_wsn}, finding an optimal window length for an effective trust computation is still an open challenge. In addition, current literatures capture the effect of time on trust computation by prioritizing recent behavior over past behavior. This approach can result in inaccurate trust assessment, potentially misclassifying an honest service provider as a malicious one and vice versa \cite{decay_function_wsn_01}, \cite{decay_function_wsn_01_improve}, \cite{decay_function_wsn_02}. Due to greater emphasis on recent trust scores in the existing works, a malicious service provider could potentially receive a higher trust value by serving only a few good transactions. Conversely, a well-known good service provider could be designated as a malicious node for a few bad transactions, which may occur due to technical issues within the system. Moreover, recent research works use computationally expensive clustering techniques to filter out malicious recommendations \cite{recommendation_clustering_02}, \cite{recommendation_clustering_01}, emphasizing the need for an efficient filtering approach to speed up the trust computation procedure.

In our research, we address the above-mentioned shortcomings and propose an efficient trust management scheme for IoT capable of ensuring better resiliency against security attacks. In particular, this paper makes the following contributions:
\begin{itemize}
\item A sliding window-based trust management scheme, where window length is adjusted dynamically based on the number of transactions inside the window to ensure necessary trust ratings for reliable trust computation. This facilitates sufficient trust scores in the window even when transactions are infrequent.  

\item A trust computation equation capturing the effect of time utilizing the harmonic mean of average trust score and average time value that balances sudden change to trust score due to the service provider's most recent behavior.

\item A recommendation filtering mechanism utilizing a modified subspace clustering algorithm \cite{CLIQUE} to rise protection against bad-mouthing and ballot-stuffing attacks. The proposed filtering mechanism executes in $\mathcal{O}(n)$ time, where $n$ is the number of points in the subspace. 

\item A rigorous security analysis of the proposed scheme demonstrating its effectiveness against on-off, ballot-stuffing, and bad-mouthing attacks. 

\item The efficacy of the proposed scheme is validated through extensive experimentation. The results indicate that the proposed scheme effectively deals with bad-mouthing and on-off attacks by precisely identifying malicious entities. It achieves approximately 44\% greater accuracy in detecting on-off attackers compared to existing methods, while performing competitively in detecting bad-mouthing attackers. Furthermore, the proposed scheme consistently identifies on-off attackers effectively, even when the percentage of these attackers increases. It also provides improved accuracy in identifying off-on attackers when multiple attacks occur simultaneously. In case of honest and malicious service providers, it demonstrates competitive performance across multiple attack scenarios. Additionally, the proposed scheme significantly  enhances efficiency by reducing the clustering time by approximately 95\%.

\end{itemize}	

The remainder of the paper is structured as follows. Section \ref{related_work} presents a literature survey on trust schemes in the IoT domain. Section \ref{proposed_scheme} presents the system model and a comprehensive description of the proposed scheme. Section \ref{security_analysis} demonstrates the robustness of the proposed scheme against different security attacks,  while Section \ref{experimental_results} presents the findings from the experimental analysis. Finally, the paper is concluded in Section \ref{sec_conclusion} with some directions for future works.

%% file: related_work.tex
\section{Related Work}\label{related_work}
In literature, different works \cite{fixed_sliding_window_01}, \cite{fixed_sliding_window_02}, \cite{fixed_window_03} utilized fixed-length sliding windows for trust management. A fixed-length sliding window is used to recode information about successful and unsuccessful transactions, and the recorded data is used to compute trust values of other devices. Gao et al. \cite{hybrid_sliding_window} employed a fixed-length sliding window and a time decay function for trust management in VANETs, where greater weights are assigned to the recent interactions. Jabeen et al.\cite{fixed_window_2021_1} for the first time considered the iteration-wise trust computation where a fixed-length window is used to record trust scores for each interation. When the window duration expires, the direct trust value of an IoT device is computed using the previous direct trust value and current window ratings. Huang et al. \cite{fixed_window_5g} implemented a trust-based intelligent collaboration interconnection system by employing a fixed-length window. This scheme uses a subjective logical framework to compute trust scores and mitigate security threats. Jiang et al. \cite{distributed_access_trust} presented an access control scheme where each IoT device maintains a fixed-length window consisting of a pre-defined number of time slots, storing the success and failure count. Sun et al. \cite{iov_recomm} proposed a hierarchical trust management framework, where vehicles evaluate direct trust and RSUs dynamically update trust assessments using a fixed-length sliding window mechanism and time decay function. When calculating direct trust for a pair of vehicles, the number of positive and negative feedback is determined by the fixed sliding window. The fixed-length window mechanisms assume that the number of transactions per time slot is constant, which is far from practicality as the number of transactions per slot can greatly vary. Frequent transactions can result in unnecessary redundant trust ratings inside the window, increasing time and memory complexity. On the other hand, infrequent transactions result in the absence of sufficient trust ratings inside the window, resulting in inaccurate trust evaluation. Hence, determining an appropriate window length by keeping a balance between the memorability of previous behavior and necessary trust scores for accurate trust computation is crucial. It is realistic to determine the window length based on the number of transactions required for accurate trust computation.

\begin{table*}
\centering
\caption{Comparison of the proposed scheme with existing works}
\label{SchemeComparisonTable}
\setlength{\arrayrulewidth}{.2mm}
\renewcommand{\arraystretch}{1.5}
\begin{tabular}{|p{2cm}|p{1 cm}|p{2cm}|p{3 cm}|p{3 cm}|p{3.5 cm}|}
\hline
\textbf{Scheme} & {\textbf{Window length}} & {\textbf{Impact of time}}  & {\textbf{Recommendation Model}} & {\textbf{Filtering Criteria}} & {\textbf{Impact of Recommendation Filtering}}  \\
\hline \hline
Jiang et al. \cite{distributed_access_trust} & Fixed & Exponential decay function & Recommendation credibility evaluation & Recommendation credibility value based on node similarity, evaluation differences, and feedback similarity & Prevents ballot-stuffing and bad-mouthing attacks\\ 
\hline
Chen et al. \cite{rte_fast_kmeans} & Fixed & Exponential decay function & Mini batch $k$-means & Collaboration records, collaborative task types and diversity of environments & Prevents ballot-stuffing, bad-mouthing, and  on-off attacks\\ 
\hline 
Sun et al.\cite{iov_recomm} & Fixed & Exponential decay function & Fuzzy C-Means algorithm & Direct trust of the recommender computed by the trust requester, similarity and difference degree between the recommending nodes against the target node & Reduces recommendation errors for bad-mouthing and ballot-stuffing attacks  \\
\hline
Khan et al. \cite{ETERS} & Dynamic (selects between two fixed lengths) & N/A & N/A & N/A & Prevents Sybil attacks, blackhole attacks, grayhole attacks, etc. \\ 
\hline
Zhang et al. \cite{time_decay_aatms} & N/A & Polynomial decay function & N/A & N/A & Prevents newcomer, on-off, collusion, etc. \\ 
\hline
Chen et al. \cite{iot_window_decay} & Fixed & Exponential decay function & $k$-means clustering & Direct trust of a recommender computed by the trust requester, similarity between the recommender and the trust requester, and confidence of the recommender & Prevents ballot-stuffing and bad-mouthing attacks\\ 
\hline
Proposed scheme & Dynamic & Weighted average and Harmonic mean & Subspace clustering & Precision Trust of IoT devices & Prevents ballot-stuffing, bad-mouthing, and on-off attacks  \\ 
\hline
\end{tabular}
\end{table*}

The dynamic sliding window mechanisms are used to overcome the limitations of the fixed-length window schemes \cite{WSN_novel, ETERS, ETAS, dynamic_window_wsn}. Khan et al. \cite{WSN_novel} proposed a trust estimation scheme using a clustering mechanism that adjusts the window length based on network scenarios. In addition, Khan et al. \cite{ETERS} proposed a trust estimation scheme that uses two fixed sliding window lengths depending on the behavior of cluster members. It selects a greater sliding length when the cluster member's behavior is good; conversely, a lower sliding length is selected when the cluster member's behavior is bad. Kumar et al. \cite{ETAS} presented a trust assessment scheme that considers communication, energy, and data trust to handle internal attacks. This scheme uses dynamic sliding window method to compute trust scores where the window length is adjusted depending on network scenarios and application requirements. Sahoo et al. \cite{dynamic_window_wsn} proposed a lightweight trust management scheme using a penalty and reward policy and a sliding window to differentiate on-off attackers and ordinary nodes. Here, the window length depends on the count of ON (misbehavior) and OFF (good behavior). Pathak et al. \cite{dynamic_window_wsn_2024} introduced a dynamically adjustable logical time window that adjusts window length according to the sensors' actions. Existing dynamic window-based trust management approaches adjust the window length based on application requirements \cite{ETAS}, network scenarios \cite{ETAS}, device limitations \cite{decay_function_wsn_01}, node actions \cite{dynamic_window_wsn_2024}, and ON and OFF count \cite{dynamic_window_wsn}. Handling numerous parameters and their complex processing when determining the window length necessitates a straightforward method for adjustment. The proposed scheme introduces a simpler mechanism to adjust the window size, where the window length is dynamically adjusted based on the number of transactions inside the window at a particular time, ensuring enough trust scores in the window for accurate trust computation even in adverse situations. 

Existing researches emphasize that the impact of experiences blurs with time and prioritize more recent trust scores to evaluate the cumulative trust scores \cite{decay_functions_social_networks, decay_functions_organizations}. Literary works use
different techniques to compute weights that are used to distinguish the past and latest trust scores. Zhang et al. \cite{decay_function_wsn_01} proposed a decay function $\theta$ to control the significance of trust scores, where $\theta$ is assigned to the current trust score while $(1-\theta)$ to the less recent trust score. In another work, Zhang et al. \cite{decay_function_wsn_01_improve} proposed a mechanism that slightly reduces the weight of the current trust score when it is high. On the contrary, if the recent trust score is low, the decay function provides more weight to the current trust value, resulting in a sharp decrease in the direct trust score. Gautam and Kumar \cite{decay_function_wsn_02} proposed a time laps function based on a forgetting curve for direct trust and a reputation function for indirect trust. Chen et al. \cite{iot_window_decay} proposed an adaptive trust model based on recommendation filtering for IoT, which uses a sliding window and time decay function for calculating the direct trust. It uses a Bayesian inference model and beta distribution to calculate the direct trust and incorporates an exponential time decay function that always gives higher weight to more recent interactions. Besides, it uses the direct trust of the recommender computed by the trust requester, the similarity between the trust requester and the recommender, and the confidence of the recommender to filter out recommendations.
Babar et al. \cite{time_decay_siot} used a polynomial equation where a decay rate constant is used to provide more weight to the recent trust score. Kong et al. \cite{time_decay_attenuation} used an exponential attenuation function to adjust the decay rate, providing more weight to the recent trust scores, where the weight of the scores decreases with time. Zhang et al. \cite{time_decay_aatms} utilized an adaptive forgetting factor to update the local trust scores of vehicles. This scheme dynamically updates the weights of current and old trust scores where the weight factor depends on the difference between the present and old scores. Jiang et al. \cite{distributed_access_trust} used Euler's co-efficient-based exponential decay function to calculate the time decay factor that again emphasizes the recent trust scores by providing the highest weight. Similarly, Chen et al. \cite{rte_fast_kmeans} deployed an exponential decay function to obtain the time decay factor that diminishes the past experiences. Wang et al. \cite{wang2025trust} proposed a dynamic trust mechanism by combining subjective and objective evaluations. The subjective evaluation calculates trust based on the end user’s direct experiences, considering both positive and negative feedback over time. This is accomplished by using a decay-weighted formula, where decay is an exponential function, ensuring that the influence of user feedback gradually decreases with time. Similarly, Sun et al. \cite{iov_recomm} computed the direct trust based on positive and negative feedback, where feedback is determined using an exponential time decay function, emphasizing the recent behavior. 
 In summary, existing techniques to introduce the effect of time prioritize recent behavior over past behavior. This approach can result in inaccurate trust estimation, depicting an honest service provider as malicious and vice versa. The proposed scheme deals with this limitation by introducing the harmonic mean of time and trust values that prevents drastic change of trust scores based on recent behavior. 
 
An IoT device can seek recommendations from other devices to estimate the trust of a device. Malicious recommenders generate different attacks such as bad-mouthing, ballot-stuffing, and on-off attacks. Shabut et al. \cite{recommendation_clustering_01} proposed a hierarchical clustering mechanism to avoid malicious recommenders. This scheme considers confidence, trust value deviation, and closeness to a cluster to select recommending nodes. It assigns the nodes to $K$ clusters based on the distance vector and uses the majority rule to select the best trustworthy cluster. The hierarchical clustering algorithm has a cost of $O(n^2)$ where $n$ is the number of nodes. Hence, heuristic algorithms are considered for hierarchical clustering. Sharma et al. \cite{recommendation_clustering_02} performed objective and subjective evaluations to filter out malicious nodes. The objective assessment assigns weight to the received recommendations based on the deviation from the average recommendation value, and the subjective evaluation ranks the recommenders based on the age of the last interaction. Rani et al. \cite{recommendation_03}  used an artificial intelligence technique and a deviation test to filter out malicious nodes, causing ballot-stuffing and bad-mouthing attacks. Jiang et al. \cite{distributed_access_trust} utilized the recommendation credibility of a node to eliminate malicious recommenders. The node similarity, evaluation difference, and feedback similarity parameters determine the recommendation credibility value. Alnasser et al. \cite{recommendation_04} prevented recommendation attacks such as bad-mouthing and ballot-stuffing attacks in VANETs using a recommendation-based trust management scheme. This mechanism equally emphasizes current and past trust values to identify malicious behavior. Moreover, this scheme uses confidence value and adaptive weight to evaluate indirect trust from recommendations to filter out harmful recommendations. Sun et al. \cite{iov_recomm} computed a reliability score based on similarity and difference degree between the recommenders, and the direct trust of the recommender calculated by the trust requester to enhance the filtering process. Then, it uses Fuzzy C-Means (FCM) algorithm to eliminate malicious recommendations, which has a complexity of $O(ndc^2t)$, where $n$ denotes the number of data points, $d$ represents the dimensionality of data points, $c$ is the number of clusters, and $t$ refers to the number of iterations required for the algorithm to converge \cite{fcm}. Chen et al. \cite{iot_window_decay} computed the trust score of a service provider using a weighted average of direct trust and recommendation trust. A trusted third party (TTP) eliminates malicious recommendations using a $k$-means clustering technique where $k$-means has a polynomial time approximation scheme of complexity 
$O(n(logn)^k\epsilon^{-2k^2d})$ where $n$ is the number of points, $k$ is the number of clusters and $d$ is the dimensionality of the data points \cite{kmeans_time}. Chen et al. \cite{rte_fast_kmeans} eliminated malicious recommendations from a cluster using the mini-batch $k$-means algorithm. The filtering process depends on collaboration records, types of collaborative tasks, and the diversity of environments. The proposed optimized mini batch $k$-means reduces the indirect trust calculation time by considering mini sets of data instead of the entire dataset. In summary, existing literature uses different clustering algorithms to filter out malicious recommendations that are time-consuming for larger data sets. The proposed scheme addresses this limitation using a customized subspace clustering method that has a time complexity of $O(n)$. In this approach, recommendations are eliminated based on the precision of IoT devices in evaluating trust scores. We present a comparative analysis of the proposed scheme against the current research works in Table \ref{SchemeComparisonTable}.

%% file: proposed_scheme.tex
\section{Proposed Scheme}\label{proposed_scheme}
\subsection{System Model}
\begin{figure}
        \centering
        
        \includegraphics[height=3 in, width = 0.45\textwidth]{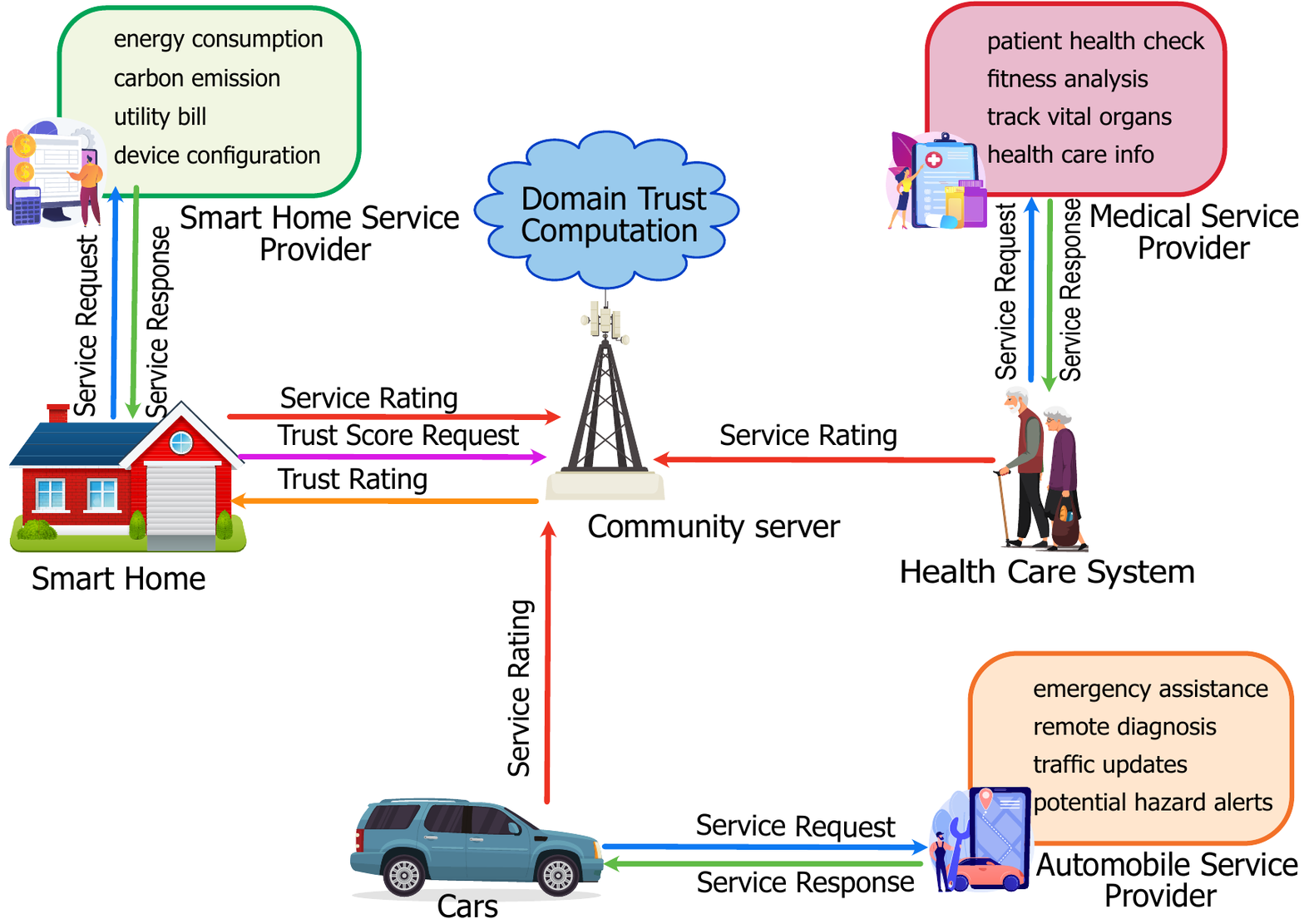}
        \caption{System model.}~\label{fig_1}
        \end{figure}

We assume that an IoT environment comprises multiple domains, where each domain is managed by a server known as \textit{community server (CS)}. Figure. \ref{fig_1} shows the system model of a single domain consisting of three key entities --- \textit{community server}, \textit{service provider}, and \textit{IoT devices} embedded into different entities.

The \textit{community server} is responsible for managing \textit{IoT devices} belonging to a particular region/domain. It computes the domain trust of a \textit{service provider} based on trust scores received from the \textit{IoT devices} in its domain. It also delivers the domain trust score of a \textit{service provider} to an \textit{IoT device} after receiving a request from that device. A \textit{service provider} offers a particular service such as GPS location \cite{gps_iot}, weather forecast \cite{forcast_service}, and computation services \cite{iot_cloud} to  \textit{IoT devices}. On the other hand, \textit{IoT devices} are low-powered sensing devices embedded into different objects around us. They collect and process information from the surrounding environment. In addition, they choose \textit{service providers} based on the domain trust scores received from a \textit{community server} and request services to carry out their operations. After receiving services, an \textit{IoT device} rates the \textit{service provider} and assigns a trust score. The \textit{community server} utilizes these trust ratings to compute the overall domain trust score of a \textit{service provider}.  

\subsection{Threat Model}
In the proposed system model, IoT devices and service providers can behave maliciously,  while the community server is considered a trusted entity. Malicious service providers engage in \textit{on-off} attacks, alternating between {ON} and {OFF} periods to disrupt reliable network operations. During the {ON} phase, the malicious service provider delivers poor-quality services, while in the OFF phase, it provides high-quality services. On the other hand, malicious IoT devices perform \textit{bad-mouthing} and \textit{ballot-stuffing} attacks.  In \textit{bad-mouthing} attacks, malicious IoT devices provide unfairly low trust ratings to honest service providers to damage their reputations. On the contrary, during \textit{ballot-stuffing} attacks,  malicious IoT devices give excessively high trust ratings to malicious service providers to inflate their trust scores. Initially, when an IoT device joins the network, it operates as an honest entity. It requests domain trust scores of service providers from the community server, receives their services, and then rates the service providers for received services. However, over time, if the device becomes compromised by adversaries \cite{IoT_node_compromise}, it may become malicious and engage in \textit{bad-mouthing} and \textit{ballot-stuffing} attacks.

\subsection{Overview of the Proposed Scheme}
An IoT device receives services from a service provider based on its requirements and stores a trust rating for each received service in its storage. It uses a sliding window-based mechanism to store and manage trust ratings, where the window length is adjusted based on the required number of transactions for accurate trust assessment. Every IoT device sends an aggregated trust score for each service provider to the community server after a predefined time interval $\delta_T$. After collecting the trust scores of a service provider from IoT devices, the community server filters out malicious trust scores by forming clusters and computes the domain trust of a service provider using the selected reliable trust scores. The notations used to describe the proposed scheme are listed in Table \ref{table:1}.

\begin{table}
\centering
\renewcommand{\arraystretch}{1.2}
\caption{List of Notation}
\label{table:1}
\scalebox{0.93}{
\begin{tabular}{ | c | l | } 
 \hline
 \textbf{Notation} & \textbf{Description} \\ [0.5ex] 
 \hline
 $Dev_i $ & $i$th IoT device\\ 
 \hline
 $SP_j$ & $j$th service provider \\
 \hline
 $\Delta_t$ & Duration of a time slot \\
 \hline
 $\delta_T$ & Time interval when $CS$ requests direct trust\\
 \hline
 $Max_{Rating}$ & Maximum number of trust ratings for trust computation\\
 \hline
  $Min_{Rating}$ & Minimum number of trust ratings for trust computation\\
 \hline
 $T_{tr}$ & Trust score factor\\ 
 \hline
 $W_T$ & Time weight factor \\  
 \hline
 $T_{intermediate}$ & Intermediate trust score computed by an IoT device  \\ [1ex] 
 \hline
 $DT_{y, x}$ & Direct trust of a service provider $y$ computed by device $x$\\ 
 \hline
 $R$ & Reward factor\\ 
 \hline
 $E$ & Penalty factor \\  
 \hline
  $TM$ & Two-dimensional direct trust matrix \\  
 \hline
 $PT$ & Two-dimensional precision matrix \\ 
 \hline
 $DTrust_y$ & Domain trust score of a service provider $y$ \\ 
 
 \hline
\end{tabular}}
\end{table}

\subsection{Trust Computation by an IoT Device}
An IoT device evaluates the delay of received packets to rate a service obtained from a service provider \cite{delay_trust}. An IoT device records the service request and service receive times and takes the differences between the two to compute a trust score. A lesser delay leads to a higher trust rating, whereas a higher delay leads to a low one. Note that trust score can be computed based on various factors such as data integrity, service reliability, energy consumption, and communication overhead \cite{ETAS}, \cite{alhandi_2023}. We plan to incorporate trust computation based on multiple factors in our future work.

\subsection{Dynamic Sliding Window Management}
An IoT device maintains a window to preserve the trust ratings of a service provider for received services. This window comprises several time slots of fixed duration $\Delta_t$. An IoT device inserts all trust ratings for the recently requested services from a particular service provider to the \textit{current} time slot. When the \textit{current} time slot reaches its maximum length $\Delta_t$, an IoT device adds it at the end of the window and adjusts the window length. The proposed scheme creates a new time slot to store the upcoming service ratings. The new time slot now becomes the \textit{current} time slot. This process is repeated after every time quantum $\Delta_t$ elapses.

After adding the new time slot, the proposed scheme adjusts the window length by removing the oldest time slot based on $Max_{Rating}$ and $Min_{Rating}$, where $Max_{Rating}$ and $Min_{Rating}$ are the maximum and minimum number of trust ratings that should be counted for trust computation, respectively. The community server can determine these parameters for a pair of service provider and an IoT device based on the service request patterns. It also updates these parameters with the change of service request patterns. A customized and simplified Non-dominated Sorting Genetic Algorithm (NSGA-II) \cite{multiobgen3} can be used to determine $Max_{Rating}$ and $Min_{Rating}$. Suppose that an IoT device has a storage capacity of $R_t$ (resource), of which $R_u$ is used to store trust scores. The total expected number of trust scores is $N_t$ of which $N_a$ trust ratings are accurate. Let the memory (resource) utilization factor is $R_f = R_u / R_t$, trust accuracy factor is $N_f = N_a / N_t$, and the fitness function is $f(max, min) = R_f / N_f$.
The community server begins with randomly generated solutions and averages two existing solutions in the crossover phase to obtain the candidate solutions for the new iteration. This iterative process is repeated until Pareto-efficiency is achieved. The maximum value should be set to ensure adequate exploration \cite{parameter11} in the algorithm. The minimum value should be set to provide a threshold for exploitation \cite{parameter11} of the algorithm. Each time window maintains a transaction count between $Max_{Rating}$ and $Min_{Rating}$.

\begin{algorithm}
\caption{Algorithm to adjust window length}\label{alg:one}
\begin{algorithmic}[1]
\Require
\Statex $wnd$ $\gets$ window of time slots of duration $\Delta_t$
\Statex $currentTimeSlot$ $\gets$ newly added time slot
\Statex $beginTimeSlot$ $\gets$ oldest time slot of a window

\Ensure
\Statex $curr\_wnd$ $\gets$ window after length adjustment
\vspace{1em}

\State $ wnd \gets wnd.add(currentTimeSlot)$  
\State $cnt \gets wnd.TransactionCount() $  
\If{$ cnt >  Max_{rating}$}
    \State $nCnt$ $\gets$ $cnt - wnd.getCount(beginTimeSlot)$ 
\While { $nCnt \geq Min_{rating}$  $\And$  $ cnt > Max_{rating} $}
 \State $wnd.delete(beginTimeslot)$
 \State $ cnt \gets wnd.TransactionCount() $ 
 \State $nCnt \gets cnt - wnd.getCount(beginTimeSlot)$ 
 \EndWhile
 \EndIf
 \State $curr\_wnd$ $\gets$ $wnd$
 \State \Return {$curr\_wnd$}

\end{algorithmic}
\end{algorithm}

\begin{figure}
    \centering
    \includegraphics[width=0.5\textwidth]{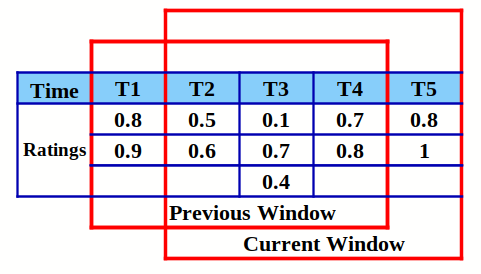}
    \caption{Propagation of the window.}~\label{Fig_2}
\end{figure}
Algorithm \ref{alg:one} describes the procedure to adjust the window length. At the end of \textit{current} time slot (when $\Delta_t$ expires),  an IoT device includes it into the existing window shown in Line 1 of Algorithm \ref{alg:one}. An IoT device also decides whether the oldest time slot should be removed from the window in line 3 $\sim$ 10. If the number of transactions ($cnt$) in the window exceeds $Max_{Rating}$, the IoT device deletes the leftmost time slot (the oldest time slot) only if the removal of the oldest time slot results in a total number of transactions ($nCnt$) greater than or equal to $Min_{rating}$ shown in line 5 $\sim$ 9. This process continues until the condition in line 5 is false.

Figure \ref{Fig_2} shows a window to manage trust ratings given to a service provider by an IoT device. For this example, we assume
$Max_{Rating}$ = 10 and $Min_{Rating}$ = 5. As shown in Fig. \ref{Fig_2}, the previous window comprises 4 time slots, containing 9 trust scores. After the expiry of $\Delta_t$, an IoT device adds $T_5$, which is the \textit{current} time slot. As the total transactions $cnt$ = 11 $> Max_{Rating}$ and the deletion of $T_1$ creates 4 time slots with $nCnt$ =9 $> Min_{Rating}$ transactions, the IoT device removes $T_1$ from the window. 

In the proposed scheme, the window size reduces with increasing rate of transactions and increases if there are less than the minimum number of trust ratings inside the window.

\subsection{Direct Trust Computation}
An IoT device computes the direct trust of a service provider using trust ratings stored in the \textit{current} window when the community server requests for the trust score at a pre-defined time interval $\delta_T$. It computes two parameters from the \textit{current} window: \textit{1) trust score factor} ($T_{tr}$) and \textit{ 2) time weight factor} ($W_T$). These parameters are used to compute $T_{intermediate}$ in Eq. \ref{Eqn_3}. The proposed scheme refers to the average value of trust scores in the \textit{current} window as trust score factor $T_{tr}$.  
\begin{equation}
 T_{tr} = \frac{1}{n}{\sum_{i=0}^{n} 
 T_{{tr}_i}}
 \label{Eqn_1}
 \end{equation}
 where $n$ is the total number of trust ratings inside the \textit{current} window, and $T_{tr_i}$ is the $i$th trust rating. An IoT device assigns a weight to each trust rating of the $j$th time slot as $W_{T_j}$ = position of the $j$th time slot/total number of time slots in the \textit{current} window.  If a time slot $T_2$ contains trust scores 0.1, 0.7, and 0.4 and it is the second time slot of the \textit{current} window comprising four time slots, then $W_{T_2}$ = 2/4 is applied to each of the trust ratings 0.1, 0.7, and 0.4. The proposed scheme refers to the average value of the time weights associated with each trust rating inside the \textit{current} window as the time weight factor $W_{T}$. 
   
\begin{equation}
W_{T}= 
\begin{dcases}
    \frac{1}{n}{\sum_{i=0}^{n} W_{{T}_i}} , & \text{if }T_{tr}\geq 0.5\\
    1-\frac{1}{n}{\sum_{i=0}^{n} W_{{T}_i}} , & \text{otherwise}
\end{dcases}
\label{Eqn_2}
\end{equation}
where $n$ is the number of trust ratings in the \textit{current} window, and $W_{T_i}$ is the time weight of the $i$th trust score. $W_{T}$ indicates the freshness of trust ratings. If $W_{T}$ is closer to 1, the trust ratings are considered from recent time slots. The lower value of $W_{T}$ indicates that the trust scores are taken from older time slots. In Eq. \ref{Eqn_2}, we consider two different time weight factors depending on whether the trust value average is less than 0.5 to diminish the effect of past experiences with time in Eq. \ref{Eqn_3}. It handles four corner cases. First, it ensures that recent and high-valued trust scores retain a high value in Eq. \ref{Eqn_3}. Second, $1-1/n\sum_{i=0}^{n}W_{T_i}$ in Eq. \ref{Eqn_2} confirms that recent and low-valued trust scores result in low values. Third, Eq. \ref{Eqn_2} ensures that past and high-valued trust scores gradually decay towards 0.5 in Eq. \ref{Eqn_3}. Finally, $1-1/n\sum_{i=0}^{n}W_{T_i}$ in Eq. \ref{Eqn_2} also guarantees that past and low-valued trust scores should approach 0.5 in Eq. \ref{Eqn_3}. 

An IoT device computes an intermediate trust value $T_{intermediate}$ using $T_{tr}$ and $W_{T}$ as shown in Eq. \ref{Eqn_3}. This equation is inspired by the $F_{\beta}$ score \cite{F1_example_02}, where $F_{\beta}$ is the harmonic mean of precision and recall. It assigns weights to two parameters based on their importance.  

\begin{equation}
T_{intermediate} = \frac{ \left({1 + \beta ^{2}}\right) \times {W_{T}} \times {T_{tr}} }{ 
\beta ^{2}\times {W_{T}} + {T_{tr}} } 
\label{Eqn_3}
\end{equation}

In Eq. \ref{Eqn_3}, the $\beta$ factor regulates the decay of effects. In a trust management scenario, an unknown and new entity is considered as uncertain. After transactions with that entity, a notion of trust is formed depending on the transactions. This notion again decays into uncertainty with time. We incorporate such an effect using the mathematical equations Eq. \ref{Eqn_2} and Eq. \ref{Eqn_3}, and the value of $\beta$ is chosen to keep the decay in the uncertain zone. We divide the trust space into three regions: (0 - 0.3) is an untrusted zone, (0.3 - 0.7) is an uncertain zone, and (0.7 - 1) is the trusted zone \cite{fixed_window_03}. $\beta$ simply reduces the impact and at most reaches uncertainty, a trust value of 0.5 (midpoint of the uncertain zone) in our scheme.

\begin{figure}    
\centering
    \includegraphics[width=0.45\textwidth]{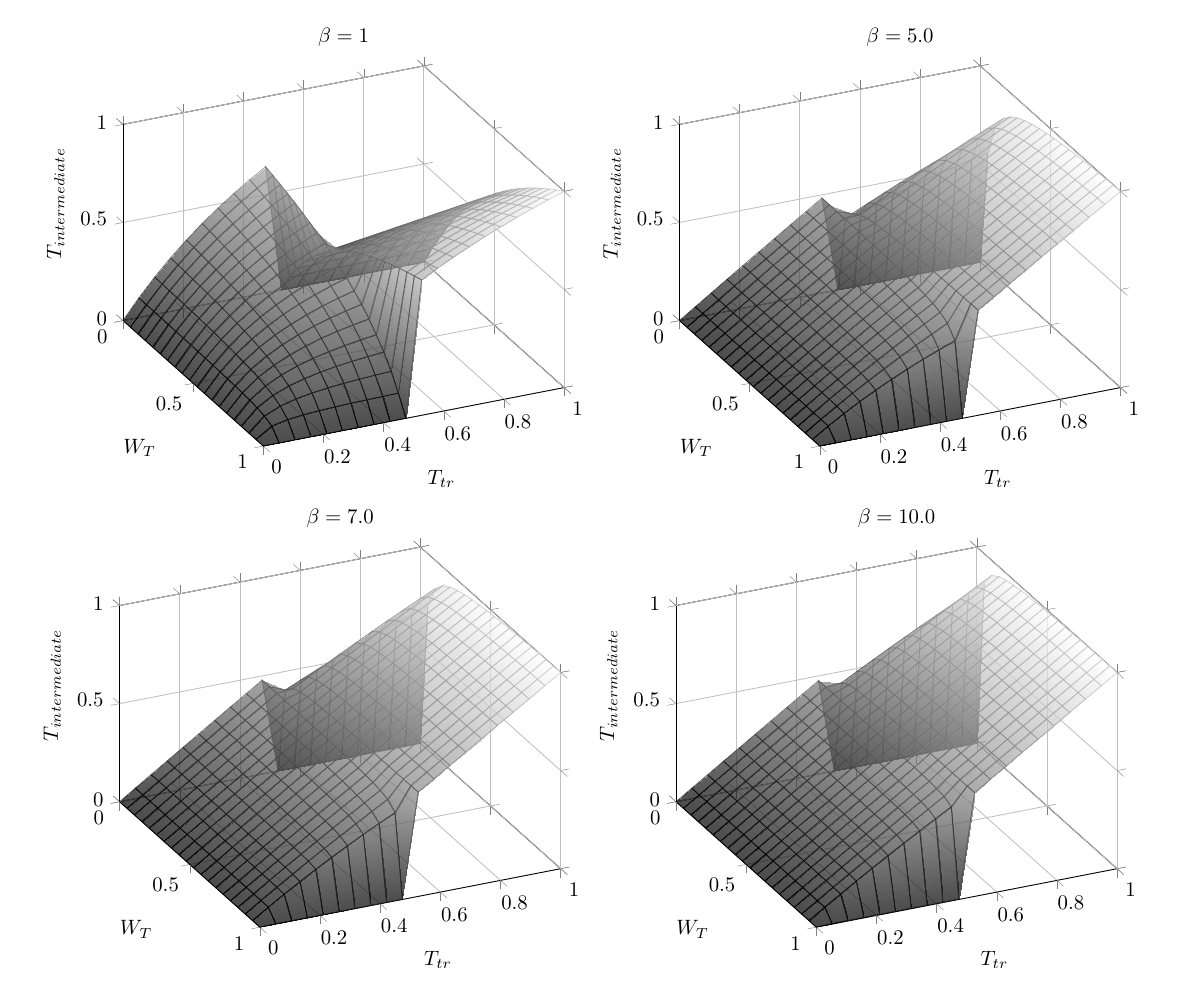}
   \caption{Effect of $\beta$ on $T_{intermediate}$.}
  \label{Fig_3}
 \end{figure}

The integration of Eq. \ref{Eqn_1}, Eq. \ref{Eqn_2}, and Eq. \ref{Eqn_3} is a form of mathematical modeling that allows the incorporation of time decay effect which diminishes the effect of experiences with time by decaying towards 0.5 by using $T_{tr}$ and $W_{T}$. This approach prevents drastic change of trust score for recent behavior, mitigating the shortcomings of existing time decay functions that prioritize recent behavior for trust computation. Figure \ref{Fig_3} illustrates the impact of $\beta$ on $T_{intermediate}$ for different values of $T_r$ and $W_T$ when $\beta$ = ${1,5,7,10}$. When $\beta$ increases, the weight of $T_{tr}$ escalates, resulting into $T_{intermediate}$ with greater reflection of $T_{tr}$. This leads to a slower rate of decay. Conversely, a decrease in $\beta$ increases the weight of $W_{T}$, which enhances the time decay effect.
Fig. \ref{Fig_3} indicates that at $\beta =7$ the trust score decays to 0.5 for the oldest ratings and does not fall below this point.

\begin{figure}
    \centering
    \subfigure[Effect of $r$ on $R$.] 
    {\includegraphics[width=.23\textwidth]{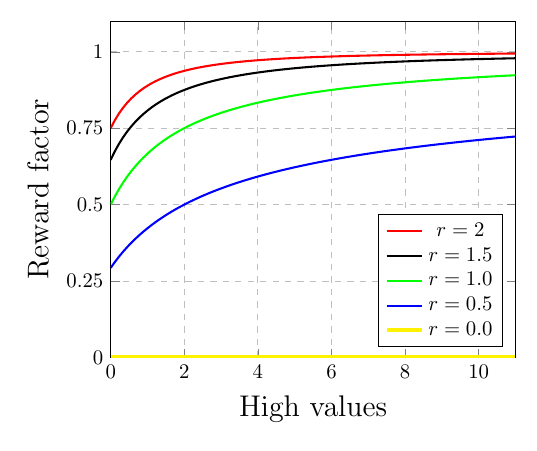}
    \label{Fig_4a}}
    \hfil
    \subfigure[Effect of $e$ on $E$.]{\includegraphics[width=.23\textwidth]{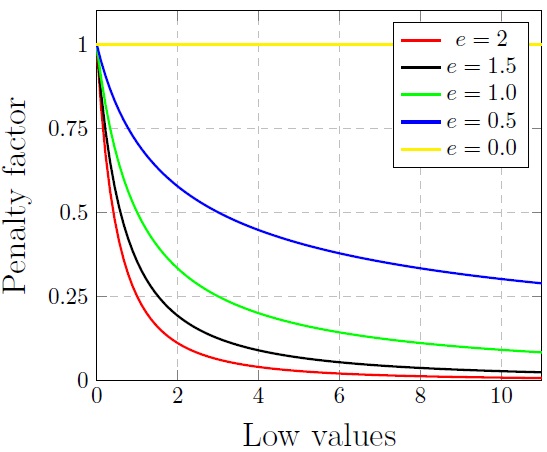}
    \label{Fig_4b}}
    \caption{Effect of $r$ and $e$.
    }
    %\vspace{-1.2em}
    \label{fig_4}
\end{figure}

Finally, an IoT device $x$ computes the direct trust $DT_{y, x}$ of a service provider $y$ using Eq. \ref{Eqn_4}. 
\begin{equation}
  DT_{y, x} = R \times E \times  T_{intermediate}
  \label{Eqn_4}
  \end{equation}

where, $R$ = $1 - \frac{1}{ (highvalues + 2)^r }$ is a reward factor and  $E$ = $\frac{1}{(lowvalues+1)^e}$ is a penalty factor. Parameters $r$ and $e$ determine the system's sensitivity towards accurate transactions and malicious transactions, respectively. Parameter $r$ can be adjusted to increase and decrease the precision with which the system differentiates between good service providers. Figure \ref{Fig_4a} shows that higher values of $r$ shift the curve to enable a service provider to get more $R$ quickly. Similarly, $e$ can be adjusted to increase and decrease the preciseness of distinguishing between malicious service providers. Figure \ref{Fig_4b} shows that a higher value of $e$ gives the graph a stepper descent, while the reduced value of e minimizes the penalty. Our scheme considers $r$ and $e$ between 0 and 2  \cite{ETERS}, \cite{Pathak2024}. An IoT device classifies a trust rating as high if the value is $>$ 0.7 and low if the value is $<$ 0.3. An IoT device keeps track of the number of high trust ratings and low ratings using \textit{highvalues} and \textit{lowvalues}, respectively. The \textit{reward} factor $R$ sets the rate at which $DT_{y, x}$ increases and it increases with the higher values of $highvalues$ and $r$. It differentiates between two service providers having the same $T_{intermediate}$ with different $highvalues$. On the other hand, the \textit{penalty} factor $E$ sets the rate at which $DT_{y, x}$ decreases. Higher values of $lowvalues$ and $e$ reduce $E$.

\subsection{Domain Trust Computation}
\begin{figure}
    \centering
    \includegraphics[width=0.35\textwidth]{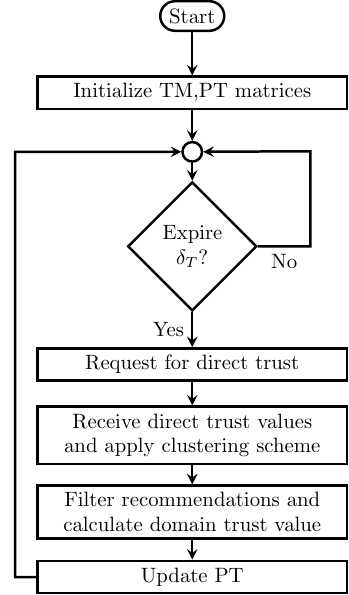}
    \caption{Workflow of the community server.}~\label{Fig_5}
\end{figure}
\begin{figure}
    \centering
    \includegraphics[width=0.45\textwidth]{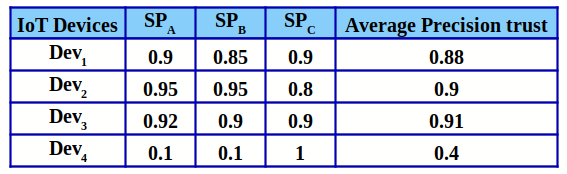}
    \caption{Precision trust matrix of a community server.}~\label{Fig_6}
    \end{figure}

In this phase, the community server computes the domain trust of each service provider following the workflow shown in Fig. \ref{Fig_5}. It requests the direct trust values of service providers from each IoT device after a pre-defined time interval $\delta_T$. Each IoT device sends back a response $DT_{y,x}$, where device $x$ rates a service provider $y$. The community server stores the collected trust values into a two-dimensional matrix $TM$. Additionally, it keeps track of the preciseness of an IoT device $x$ in computing the trust value of a service provider $y$ in a two-dimensional precision matrix $PT$. Initially, all entries in this matrix are set to 1, based on the assumption that each service is precisely evaluated by all end devices, which may deviate over time. In successive iterations, the community server uses the $PT$ refined in the earlier iteration and employs a clustering algorithm to classify the IoT devices (trust raters) into several groups. This classification filters out the malicious trust raters, and the community server computes the domain trust using the direct trust provided by the trustworthy trust raters. Finally, the community server updates the $PT$ for the next iteration of computing the domain trust. Figure \ref{Fig_6} shows a precision matrix where four IoT devices evaluate three service providers. Here to note that the average precision trust of $Dev_1$ is the average of the precision trust of the device in judging service providers $SP_A$, $SP_B$, and $SP_C$.

\subsubsection{Cluster Computation}
The community server assigns each IoT device into one of the following three categories based on its average precision trust value: 
\begin{itemize}

\item Honest trust rater (average precision trust $>$ 0.7)

\item Uncertain trust rater (average precision trust in 0.3 $\sim$ 0.7)

\item Malicious trust rater (average precision trust $<$ 0.3)   
\end{itemize}

\begin{algorithm}
\caption{Algorithm for cluster formation }\label{alg:two}
\begin{algorithmic}[1]
\Require 
\Statex $TM$: Direct trust matrix 
\Statex $PT$: Precision trust matrix 
\Statex $SP\_ID$: Service provider's ID
\Statex $Grids[3]$: List of three vectors
\Statex $Avg\_Prec[3]$: Average precision trust of vectors in $Grid [3]$ 
\Statex $isDense[3]$: Boolean values representing whether a grid is dense or not 
\Statex $totalPoints$: Number of IoT devices
\Statex $ actualCluster$: Number of actual clusters   

\vspace {1.5em}
\State $minPoints \gets totalPoints/3$
\Comment{A dense grid holds at least one-third number of points} 
\For{$i\leftarrow 1$ to  $totalPoints$}

\State $pt.x  \gets TM[SP\_ID][i] $  \Comment{direct trust assigned by $i$}
\State $pt.y  \gets PT[SP\_ID][i] $ \Comment{precision trust of $i$}
\If{ $pt.x  \geq 0.7 $  }
\State $ Grids[2].push(pt)$ \Comment{Add to the rightmost grid} 
\State $pt.cluster \gets 2$

\ElsIf { $pt.x  \geq 0.3 $  }
\State       $ Grids[1].push(pt)$ \Comment{Add to the central grid}   
\State $pt.cluster \gets 1$

\Else 
\State $ Grids[0].push(pt)$ \Comment{Add to the leftmost grid}
\State $pt.cluster \gets 0$
\EndIf
\EndFor

\State $maxAvgPrec \gets -\infty$
\For{$j\leftarrow 0$ to $2$}
\State $count \gets Grids[j].getcount()$
\If{$count \geq minPoints$} 
  \State $isDense[j] \gets true$   
  \State $Avg\_Prec[j] \gets grids[j].getavgprecision()$

  \If{$Avg\_Prec[j]  \geq maxAvgPrec$}
  \State $maxAvgPrec \gets Avg\_Prec[j]$
  \State $actualcluster \gets j $ 
  \EndIf
  \Else
  \State $isDense[j] \gets false $
  \EndIf
\EndFor
\end{algorithmic}
\end{algorithm}

\begin{figure}
        \centering
        \includegraphics[width=0.45\textwidth]{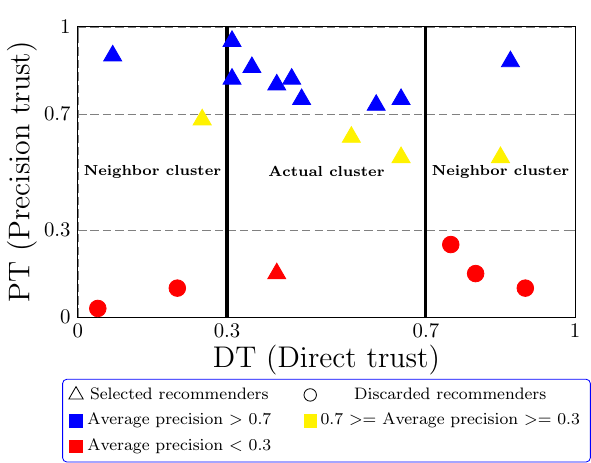}
        \caption{Clustering IoT devices.}~\label{Fig_7}
    \end{figure}

After collecting direct trust scores, the $CS$ also divides IoT devices into three clusters based on the received direct trust and precision trust. Our scheme uses 0.3 and 0.7 to divide the data space into three clusters \cite{fixed_window_03}. It uses a subspace clustering algorithm \cite{CLIQUE} to form clusters, which is modified to consider a one-dimensional data space and to identify the dense subspaces. The overall process of forming clusters is described in Algorithm \ref{alg:two}. For a particular service provider $SP\_ID$, the community server assigns an IoT device into a grid/cluster based on the direct trust it provides to $SP\_ID$ (discussed in lines 2 $\sim$ 15). After that, the community server finds a dense region in each cluster in line 17 $\sim$ 29. A dense region contains at least one-third number of data points. For a dense region, the cluster manager computes the average precision trust score of the IoT devices belonging to that region and selects a cluster having maximum average precision trust score as an \textit{actual cluster}. If there is a tie, the rightmost cluster is designated as the \textit{actual cluster}. Immediate adjacent clusters of the \textit{actual cluster} are known as the \textit{neighbor clusters}. The remaining cluster is designated as the \textit{wrong cluster}. Figure. \ref{Fig_7} shows clusters formed using Algorithm \ref{alg:two}, where the x-axis and y-axis represent the direct trust and average precision trust scores, respectively. Here, region $0.3 \sim 0.7$ (x-axis) is the \textit{actual} cluster. The leftmost \textit{neighbor} cluster ($0\sim0.3$ region) contains IoT devices that provide low direct trust scores, and the IoT devices in the rightmost \textit{neighbor} cluster ($0.3 \sim 0.7$ region) give high direct trust scores to a service provider.

\subsubsection{Filtering Malicious Trust Values}\label{filtering_recommendations}
After constructing the clusters, the community server considers all direct trust scores that satisfy the following conditions:  

\begin{enumerate}
\item Trust ratings in the actual cluster.

\item Trust scores from the neighbor cluster given by IoT devices having an average precision trust $>$ 0.3. Due to the deviation of trust values from the actual cluster's scores, trust scores from the neighbor cluster will only be considered if they are provided by an IoT device with a history of correctly rating the service. 

\item Trust scores from the wrong cluster provided by IoT devices having an average precision trust rating $>$ 0.7. Due to their higher deviation from the actual cluster's ratings, trust scores from the wrong cluster will only be considered if they are given by IoT devices having a very good history of correctly rating services.  
\end{enumerate}
Trust ratings that do not satisfy the above-mentioned conditions are not considered for domain trust computation of a service provider. 

\subsubsection{Domain Trust Computation}
The community server computes the domain trust of a service provider $y$ as follows:
\begin{equation}
\label{Eqn_5}
   DTrust_{y} = 0.5 \times (DTrust_{y_{previous}} + DTrust_{y_{new}})
\end{equation}
where, $DTrust_{y_{previous}}$ and $DTrust_{y_{new}}$ are the domain trusts of the previous iteration and current iteration, respectively. $DTrust_{y_{new}}$ is computed as the average of direct trust scores selected by filtering malicious trust scores discussed in Section \ref{filtering_recommendations}. The $CS$ assigns equal weight to both the domain trust of the previous iteration and the domain trust of the current iteration in computing the domain trust to resist the on-off attack by preventing sudden drastic change of trust values. The idea of using previous trust scores in trust computation is considered in different literary works \cite{recommendation_04}, \cite{tarf}. Providing more weight to previous trust scores makes the system slower in changing trust ratings, i.e., the system becomes less reactive. This would allow occasional faults of good service providers to be less costly. But, it allows the attackers more opportunities to exploit the system. On the contrary, giving more weight to recent scores makes the system more reactive, leading to quick detection of attackers. Nevertheless, it increases the probability of good service providers with occasional technical faults being identified as malicious. The proposed scheme obtains a good balance of this trade-off by providing equal importance to the past and present domain trust scores (Eq. \ref{Eqn_5}), and by using harmonic mean of average trust scores and average time values (Eq. \ref{Eqn_1}, Eq. \ref{Eqn_2}, and Eq. \ref{Eqn_3}), preventing sudden trust fluctuations due to recent behavior.

\subsubsection{Update Precision Trust Matrix}
After computing the domain trust, the community server updates the precision trust matrix. If the direct trust score $DT_{y, x}$ of a service provider $y$ given by an IoT device $x$ is inside the actual cluster then the precision matrix entry $PT_{y, x}$ is updated as follows:

\begin{equation}
\label{eq_6}
PT_{y,x} = \frac{1}{2} \times ( PT_{y,x} + 1 ) 
\end{equation}

\noindent This increases $PT_{y,x}$ towards a trust score of 1. An IoT device $x$ correctly rates a service provider $y$ when the direct trust is inside the actual cluster. Hence, the precision trust of $x$ for evaluating $y$ is rewarded. 

If $DT_{y, x}$ is inside the neighbor cluster, then $PT_{y, x}$ is updated as follows:

\begin{equation}
\label{eq_7}
PT_{y,x} = \frac{1}{2} \times ( PT_{y,x} + \frac{1}{2} )
\end{equation}

\noindent This equation increases $PT_{y,x}$ if it is less than 0.5, where 0.5 is the middle point of the trust range $(0 \sim 1)$ and the trust score of an unknown entity. On the other hand, this equation decreases $PT_{y,x}$ if it is higher than 0.5. 
For the first case, the precision trust value increases. We assume that a device having a precision trust $<$ 0.5 indicates that previously it was inside the wrong cluster, and its precision is gradually increasing. As now it is more precise in rating, Eq. \ref{eq_7} increases its precision trust value. 
For the second case, a device with a precision trust $>$ 0.5 gets a reduction. We assume that it had been more precise before. Hence, Eq. \ref{eq_7} changes its precision trust to a lower value due to its recent deviation from the actual cluster.

If $DT_{y, x}$ belongs to the wrong cluster, the precision matrix entry $PT_{y, x}$ is updated as follows:
\begin{equation}
\label{eq_8}
PT_{y,x} = \frac{1}{2} \times ( PT_{y,x} + 0 )
\end{equation}
\noindent This equation decreases $PT_{y,x}$ toward 0. As the trust value is in the wrong cluster, the IoT device $x$ shows poor performance in computing a trust value and is expected to show the same behavior in the future. Hence, $x$ receives a reduction in the precision trust score.

%% file: security_analysis.tex
\section{Security Analysis}\label{security_analysis}
In this section, we discuss the proposed scheme's resiliency against on-off, bad-mouthing and ballot-stuffing
attacks.

\subsection{On-Off Attacks}\label{on-off-security}
Each IoT device uses a dynamic sliding window to preserve trust scores of a service provider where the length of the window changes depending on the necessary number of transactions required to compute a reliable trust score.
Unlike the previous schemes, our scheme tries to contain at least the minimum number of ratings needed for trust computation in the window. This allows a device to have more opportunity to check
if the service provider is an on-off attacker or not.
    \begin{figure}
    \centering
    \includegraphics[width=0.85\linewidth]{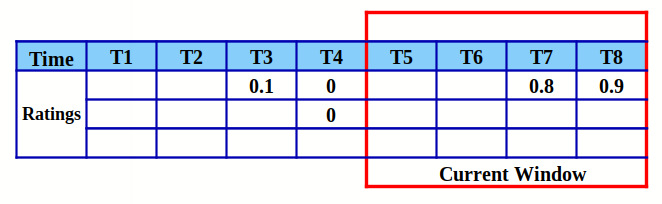}
    \caption{Fixed-length window problem.}~\label{Fig_8}
    \end{figure}
Figure. \ref{Fig_8} shows a window that always maintains a fixed number of time slots. As this window maintains a fixed number of time slots while sliding the window, the IoT device is currently unaware of previous bad ratings. Hence, the device treats the service provider as honest while it is actually performing the on-off attack.
\begin{figure}
    \centering
    \includegraphics[width=0.45\textwidth]{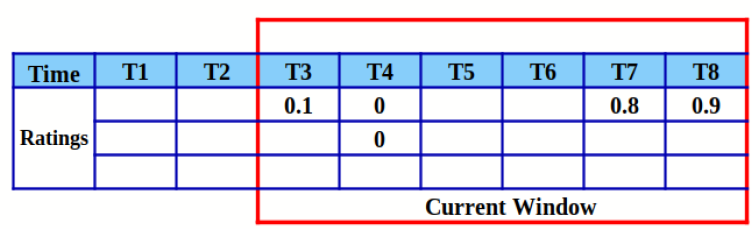} %fig_11.pdf
    \caption{Dynamic-length window. }~\label{Fig_9}
    \end{figure}
Figure \ref{Fig_9} shows a dynamic-length window that follows the proposed scheme to adjust the length. The window size increases to keep the minimum number of ratings required to compute a trust score. The proposed scheme allows the device to remember previous bad ratings, which increases the chance of the device to correctly detect an attacker.
  
Besides, the proposed scheme multiplies $T_{intermediate}$ with the reward factor $R$ and penalty factor $E$ in Eq. \ref{Eqn_4}. The penalty factor $E$ adjusts $T_{intermediate}$ depending on the number of bad transactions, which reduces the scope of attackers to execute on-off attacks. For example, an IoT device computes $T_{intermediate}$ = 0.6 for a service provider A with \textit{highvalues} = 10 and \textit{lowvalues} = 10. The same IoT device computes $T_{intermediate}$ = 0.6 for a service provider B with \textit{highvalues} = 19 and \textit{lowvalues} = 1. From Eq. \ref{Eqn_4}, it is clearly visible that the IoT device yields a lower direct trust score for service provider A than service provider B. Clearly, service provider A is executing the on-off attack, while service provider B consistently maintains its service quality. Hence, the direct trust score in Eq. \ref{Eqn_4} reflects the effect of on-off attacks.

The community server uses the current and previous domain trust scores to compute the final domain trust of a service provider using Eq. \ref{Eqn_5}, where it assigns equal weight to both trust scores. If the domain trust score is computed using the present score only, then a malicious service provider can reach from 0 to 1 in one iteration, giving the on-off attackers more opportunity to quickly reduce and increase the service quality. The previous trust value in Eq. \ref{Eqn_5} prevents the rapid fluctuation of trust scores, preventing the attackers from executing on-off attacks frequently. 

\subsection{Bad-Mouthing Attacks}
The proposed scheme utilizes a customized subspace clustering method to filter out poor recommendations. The community server identifies an actual cluster comprising IoT devices working with high precision using this clustering mechanism and finally selects recommender nodes using the rules discussed in Section \ref{filtering_recommendations}. In bad-mouthing attacks, malicious IoT devices having low precision trust scores intentionally provide low trust ratings to an honest service provider. As the proposed scheme always chooses an actual cluster with high average precision trust and selects reliable nodes from other clusters, our scheme always selects reliable IoT devices as recommender nodes, filtering out the low-precision malicious IoT devices. Hence, our scheme effectively handles bad-mouthing attacks.
\begin{figure}
    \centering
    \includegraphics[width=.45\textwidth]{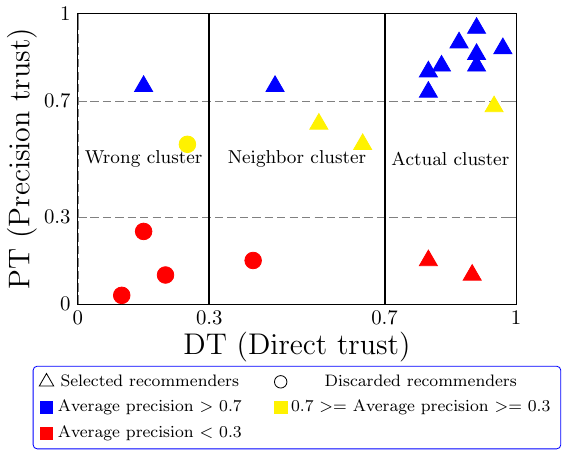}
    \caption{Recommendations during bad-mouthing attacks.}
    \label{Fig_10}
\end{figure}

Figure \ref{Fig_10} shows the filtering process for bad-mouthing attacks. Here, the blue nodes have an average precision trust score of $>$ 0.7, the yellow nodes have an average precision trust of $>$ 0.3, and the red nodes have an average precision trust of $<$ 0.3. Triangle nodes are the ones whose recommendations are selected, and the circular nodes are filtered out from the domain trust calculation procedure. As shown in Fig. \ref{Fig_10}, after the clustering process, the community server selects the rightmost cluster as the actual cluster. All the points marked as triangular are selected for domain trust computation (using rule 1 of Section \ref{filtering_recommendations}). From the neighbor cluster CS only selects the blue and yellow triangular nodes (using rule 2 of Section \ref{filtering_recommendations}). From the wrong cluster, only blue nodes are selected (using rule 3 of Section \ref{filtering_recommendations}). The rest of the nodes marked as circular are filtered out. In this way, the community server reduces the impact of bad-mouthing attacks by eliminating IoT devices with low precision trust. 

\subsection{Ballot-Stuffing Attacks}
In ballot-stuffing attacks, malicious IoT devices (with low precision trust) give high trust scores to a malicious service provider. As the proposed scheme emphasizes the precision trust scores in selecting recommender nodes, it avoids low-precision IoT devices providing high direct scores from the domain trust computation. Hence, the proposed scheme handles ballot-stuffing attacks effectively.

\begin{figure}
    \centering
     \includegraphics[width=.45\textwidth]{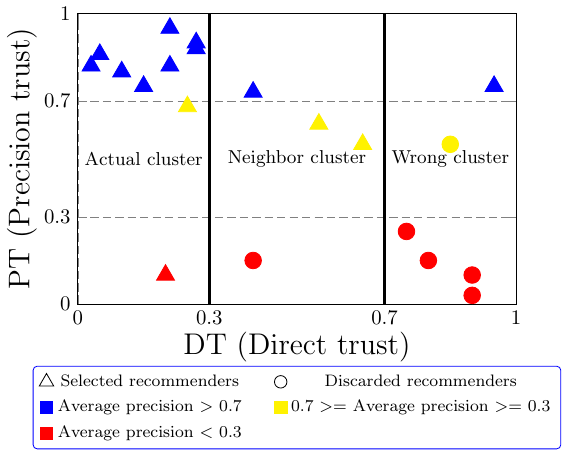}
       \caption{Recommendations during ballot-stuffing attacks.}
      \label{Fig_11}
\end{figure}

Figure \ref{Fig_11} shows the filtering process for ballot-stuffing attacks. The community server selects the leftmost cluster as the actual cluster. All the points selected from the actual cluster are marked as triangles (using rule 1 of Section \ref{filtering_recommendations}). From the neighbor cluster, the CS only selects the blue and yellow triangular nodes (using rule 2 of Section \ref{filtering_recommendations}), and from the wrong cluster only blue nodes are selected (using rule 3 of Section \ref{filtering_recommendations}). The rest of the nodes marked as circles are filtered out, and this mitigates the impact of ballot-stuffing attacks.

%% file: experimental_result.tex
\section{Experimental Evaluation}\label{experimental_results}
For the performance evaluation of the proposed scheme, we considered trust score convergence of service providers, resilience against bad-mouthing, ballot-stuffing, and on-off attacks, as well as the computation time of clusters. To demonstrate the effectiveness of the proposed scheme, its performance was compared with that of the Adaptive Trust Model based on Recommendation Filtering (ATMRF) scheme \cite{iot_window_decay}, and the Trust Model based on Recommendation Filtering (TMRF) scheme \cite{iov_recomm}.

We made significant modifications to ATMRF \cite{iot_window_decay} and TMRF \cite{iov_recomm} to have a meaningful comparison with the proposed scheme. The TMRF \cite{iov_recomm} was adapted for a static IoT environment to facilitate comparison with the proposed scheme. We opted for a single Trusted Third Party (TTP) in both schemes to maintain consistency with the community server model in the proposed scheme. Besides, a distinctive category of nodes, known as service providers, was introduced in both schemes without interrupting the operation of the remaining network. The TMRF was modified to maintain transaction histories within the IoT devices. Both schemes allow each IoT device to maintain a trust score for transactions within a time window instead of only successful (1) and unsuccessful (0) information. The domain trust of a service provider in both schemes was computed by averaging trust scores assigned to a service provider by IoT devices.  Furthermore, in TMRF, recommendation-based trust, previously calculated on demand, is now incorporated into the global trust evaluation process. 

\begin{table}
\centering
\caption{Parameters used in Simulation}
\label{simulationTable}
\setlength{\arrayrulewidth}{.2mm}
\renewcommand{\arraystretch}{1.5}
\scalebox{0.85}{
\begin{tabular}{|p{4cm}|p{5.5cm}|}
\hline
\textbf{Parameter}    & {\textbf{Value}}  \\
\hline \hline
Platform  & Ubuntu 20.04 \\
\hline
MAC and physical layer protocol & IEEE 802.11g WiFi protocol   \\ 
\hline
Trasport layer protocol & TCP\\
\hline
Network layer protocol & IP     \\
\hline
Number of IoT device & 150     \\
\hline
Number of service provider & 5\\
\hline
Number of community server &  1 \\
\hline
Time slot interval & 20 sec. \\
\hline
Domain trust calculation interval & 100 sec.\\
\hline
Simulation time & 5000 sec.       \\
\hline
Mobility & N/A \\
\hline
$\beta$ & 7.0  \\
\hline
$Max_{Rating}$ & 20  \\
\hline
$Min_{Rating}$ & 5  \\
\hline
$r$ & 1.5 \\
\hline
$e$ & 0.25 \\
\hline
\end{tabular}}
\end{table}
\begin{figure*}
    \centering
    \subfigure[Honest service provider.] {\includegraphics[width=.40\textwidth]{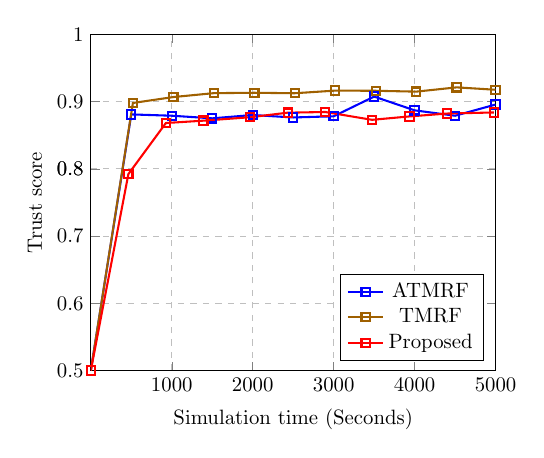}
    \label{fig_12a}}
    \hfil
    \subfigure[Malicious service provider.]{\includegraphics[width=.40\textwidth]{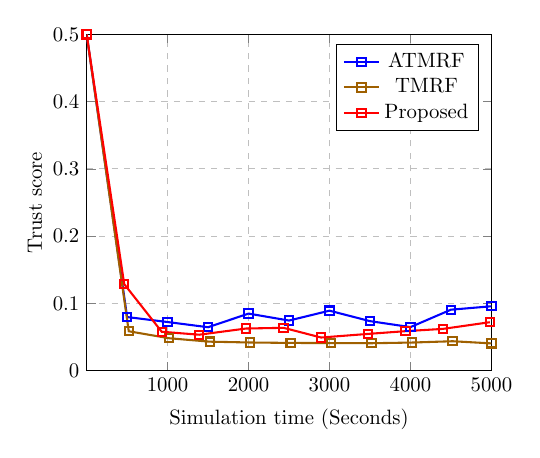}
    \label{fig_12b}}
    \hfil
    \caption{Trust value convergence of a service provider.}
    \vspace{-1.2em}
    \label{fig_12}
\end{figure*}
We implemented all the schemes using a discrete event network simulator NS-3 \cite{riley2010ns}. We developed code modules for IoT devices, community servers, and service providers and attached them to the corresponding NS-3 nodes. WiFi network protocol was chosen as the communication protocol among the NS-3 nodes. The simulation parameters are shown in Table \ref{simulationTable}. 
In our experiments, the time window holds four to five time slots \cite{ETERS}, \cite{iot_window_decay}. Hence, the minimum time slot average can be $W_{T} = 1/5$. We chose an arbitrary maximum decay of less than 10\% of the actual trust score average for the experiments. To obtain this decay rate we set $\beta = 7.0$, which is used to transform the earliest honest or malicious trust values into uncertain trust values. $Max_{Rating}$ was chosen to ensure five time slots in a time window \cite{ETERS}, \cite{iot_window_decay}. $Min_{Rating}$ ensures a minimum of one transaction in each time slot \cite{ETERS}, \cite{iot_window_decay}. $r$ = 1.5 ensures two to three high ratings are required to receive a reward factor of 1.0, and $e$ = 0.25 requires at least 10-15 low ratings to receive an error factor of 0.25. We chose this combination of $r$ and $e$ from experimental evaluations as these values provide most accurate trust scores aligning with the simulation scenarios. Both $r$ and $e$ can be adjusted based on overall service quality. 

In all the schemes, the community server set the trust value of a service provider to 0.5. After every four seconds, nodes chose a service provider, took services, computed a trust rating for the received services, and inserted trust ratings into a time window. We adjusted the time window by adding a time slot after every 20 seconds. The community server executed the domain trust calculation process once in each iteration, where an iteration length was 100 seconds. In the proposed scheme, at the end of 100 seconds, the community server collected direct trust scores from IoT devices, computed clusters, and filtered out bad recommendations. Finally, the community server calculated the service providers' domain trust scores and updated the IoT devices' precision trust values. In ATMRF, after every 100 seconds, all IoT devices transmitted their respective time windows and requested the community server to provide the trust score for a given service provider. Subsequently, the community server computed the trust score utilizing the received time windows. In TMRF, the IoT devices regularly sent the feedback of their interactions to the community server, and after every 100 seconds, the trust score was calculated in the community server based on those reported feedback. Both ATMRF and TMRF compute trust scores using a combination of direct trust and recommendation trust. The ATMRF \cite{iot_window_decay} employs the $k$-Means algorithm, the similarity between the recommender and the trust requester, the confidence of the recommender, and the direct trust of the recommender computed by the trust requester. In comparison, the TMRF \cite{iov_recomm} utilizes the Fuzzy C-Means algorithm, considers the similarity and the difference between the recommenders, along with their direct trust relationship with the trust requester, to compute trust scores. Experimental results discussed in this section can be found in \cite{exp_data}.

\subsection{Convergence of Trust Values}
In this experiment, 150 IoT devices evaluated the services provided by five SPs. To demonstrate the trust convergence of an honest SP, all SPs behaved honestly. An honest SP delivered 95\% of the services on time and 5\% services were delayed. In contrast, to show the trust score convergence of a malicious SP, all SPs performed maliciously. Here, a malicious SP delayed 95\% of the services, while delivered 5\% on time. In both cases, the trust scores of the SPs are averaged to determine the mean trust score of a service provider.

Figure \ref{fig_12} shows the domain trust convergence of a service provider. Initially, the trust value of a service provider is 0.5, which gradually increases with the simulation time for an honest service provider as shown in Fig. \ref{fig_12a}. We observe from Fig. \ref{fig_12a} that trust scores increase rapidly in the earlier phase of the simulation, and then maintain almost a constant value for the rest of the simulation time for all three schemes. Similarly, in Fig. \ref{fig_12b} the domain trust value starts at 0.5 and gradually decreases with simulation time for a malicious service provider, where trust scores sharply decrease early in the simulation process and then maintain a constant value for the remaining simulation time. Figures \ref{fig_12a} and \ref{fig_12b} demonstrate that the TMRF performs better in identifying honest and malicious service providers, whereas the proposed scheme and ATMRF display similar performance in computing trust scores. The TMRF enhances its clustering mechanism with an additional reliability assessment for each recommender, evaluating the similarity and difference of their recommendations, along with the direct trust relationship with the trust requester, which is later used to weight trust values. The ATMRF and TMRF converge faster in the first 1000 seconds due to their reliance on clustering algorithms ($k$-Means and Fuzzy C-Means, respectively), which quickly stabilize based on direct and recommendation trust scores. The proposed scheme uses previous and present domain trust scores to compute the current trust score. As a result, it may need a few additional iterations (typically three to four) to refine the initial trust estimate and achieve stable and accurate results.

\subsection{Resiliency against Bad-Mouthing Attack}
We used the Mean Absolute Error (MAE) to demonstrate the system's resiliency against attacks, where MAE is the average difference of domain trust values from the ground truth. We conducted experiments by varying malicious IoT devices between 10\% $\sim$ 60\% and computed MAE values after every 100 seconds. Each IoT device assessed the services provided by five SPs. For bad-mouthing attacks, all SPs were honest, and for ballot-stuffing attacks, all SPs were malicious. The simulation initially started with 10\% malicious IoT devices, which increased by 10\% after every 800 seconds. For each percentage of malicious IoT devices, we recorded MAE values for every 100 seconds for each SP over 800 seconds. The overall MAE for a particular percentage of malicious IoT devices was determined by taking the mean of the MAEs of all SPs.

\begin{figure}
    \centering
    \includegraphics[width=0.40\textwidth]{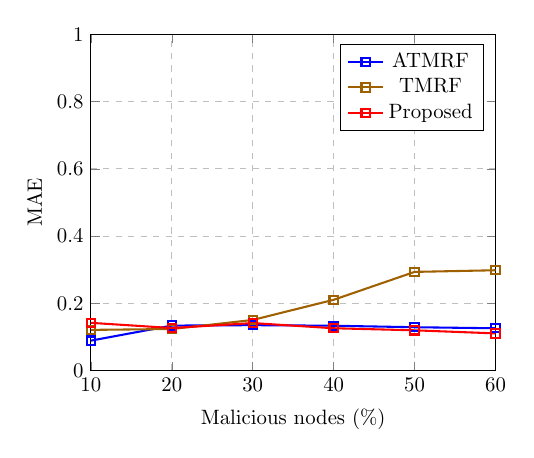}
    \caption{Resiliency against bad-mouthing attack.}
    \label{fig_13}
\end{figure}

Figure \ref{fig_13} shows that the proposed scheme and ATMRF exhibit similar MAE in most of the cases and the TMRF struggles to keep the MAE down as the percentage of malicious nodes increases. In bad-mouthing attacks, malicious IoT devices give low trust scores to an honest service provider, while honest nodes provide higher trust scores. The proposed scheme filters out the direct trust scores provided by malicious nodes and utilizes trust ratings provided by the honest nodes to compute domain trust scores of service providers. The windows of honest nodes primarily hold high trust scores ($>$ 0.7) for an honest service provider. Due to the absence of low trust scores ($<$ 0.3), $e$ (used in the penalty factor in Eq. \ref{Eqn_4}) has a negligible effect on overall trust score calculation. $r = 1.5$ is set in a manner that yields a trust score for the proposed scheme that is approximately equal to that of ATMRF. In ATMRF, utilization of $k$-Means clustering, direct trust of the recommender computed by trust requester, similarity between trust requester and recommender, and confidence of the recommender reduce the influence of bad recommendation and keeps the MAE low. TMRF’s adaptive weighting uses a hard threshold on the  average direct trust of its recommenders. As soon as a few low ratings from bad-mouthing push that average below 0.5, TMRF throws away all the recommendation trusts and falls back to using the direct trust only, causing its MAE to spike.

Figure \ref{fig_13} shows that with 10\% malicious nodes, the proposed scheme attains almost 60\% higher MAE than the ATMRF. The proposed scheme has a lower reward rate to ensure a slower increase in trust value for an honest service provider. Additionally, the proposed scheme takes a weighted average of previous and current domain trust scores. For these reasons, the proposed scheme initially takes more time to reduce the MAE than the ATMRF and TMRF schemes.

\begin{figure}
    \centering
    \includegraphics[width=0.40\textwidth]{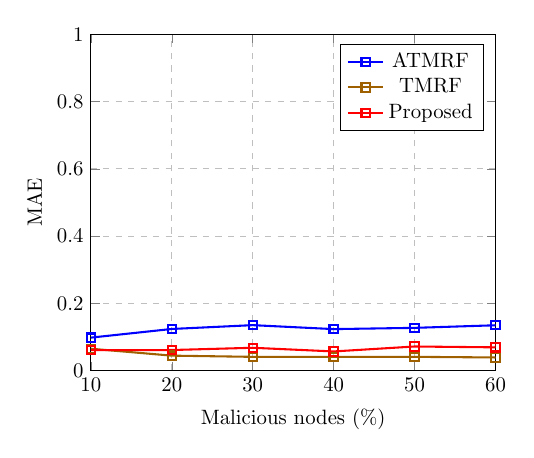}
    \caption{Resiliency against ballot-stuffing attack.}
    \label{fig_14}
\end{figure}

\subsection{Resiliency against Ballot-Stuffing Attack}

Figure \ref{fig_14} demonstrates that the proposed scheme outperforms ATMRF in most cases and shows performance similar to TMRF, where TMRF exhibits slightly better performance. The proposed scheme reduces MAE by almost 71\% compared to the ATMRF when there are 60\% malicious nodes. On average, the proposed scheme attains nearly 63\% improvement over ATMRF. In ballot-stuffing attacks, malicious IoT devices give high trust ratings to a malicious service provider, while honest IoT devices provide low trust ratings. The proposed scheme eliminates the ratings of malicious IoT devices and considers the trust ratings provided by honest IoT devices for the domain trust calculation of a service provider. The windows of honest IoT devices contain mostly low trust ratings ($<$ 0.3) for the malicious service provider, imposing a significant penalty in the proposed scheme for the presence of penalty factor $E$ in direct trust calculation (Eq. \ref{Eqn_4}). This penalty reduces the direct trust score more in our scheme in comparison to the ATMRF \cite{iot_window_decay}. On the other hand, despite showing a modest 8\% improvement over TMRF \cite{iov_recomm} at 10\% malicious nodes, the proposed scheme underperforms TMRF by approximately 36\% on average in other percentage scenarios. The adaptive weighting mechanism of TMRF prevents MAE from increasing with the growing percentage of malicious nodes in ballot-stuffing attacks. As the inflated recommendations keep the filtered average $>$ 0.5, TMRF remains in its balanced mode and continue to maintain a low MAE by combining direct and recommendation trust. 

\subsection{Resiliency against On-Off attack}\label{onoff_revision_april}

\begin{table}
\centering
\caption{On-Off attack case descriptions and results }
\label{table_4}
\setlength{\arrayrulewidth}{.2mm}
\renewcommand{\arraystretch}{1.5}
\scalebox{0.95}{
\begin{tabular}{|p{0.4cm}|p{3.2cm}|p{1.0cm}|p{1.3cm}|p{1.3cm}|}
\hline
\textbf{Case ID}    & {\textbf{On-off interval}} & {\textbf{Service request interval}} & {\textbf{Trust score reduction (\%) with respect to ATMRF}} & {\textbf{Trust score reduction (\%) with respect to TMRF}}  \\
\hline \hline
$A1$  & 30 sec. (ON) - 70 sec. (OFF) & 4 sec. & 71\% & 69\%\\
\hline
$A1^{\prime}$  & 70 sec. (OFF) - 30 sec. (ON) & 4 sec. & 66\% & 68\%\\
\hline \hline

$A2$  & 40 sec. (ON) - 60 sec. (OFF) & 4 sec. & 63\% & 61\%\\
\hline
$A2^{\prime}$  & 60 sec. (OFF) - 40 sec. (ON) & 4 sec. & 59\% &62\%\\
\hline \hline

$A3$  & 50 sec. (ON) - 50 sec. (OFF) & 4 sec. & 55\% & 54\%\\
\hline
$A3^{\prime}$  & 50 sec. (OFF) - 50 sec. (ON) & 4 sec. & 28\% & 47\% \\
\hline \hline

$A4$  & 60 sec. (ON) - 40 sec. (OFF) & 4 sec. & 41\% & 40\%\\
\hline
$A4^{\prime}$  & 40 sec. (OFF) - 60 sec. (ON) & 4 sec. & 31\% & 38\%\\
\hline \hline

$A5$  & 70 sec. (ON) - 30 sec. (OFF) & 4 sec. & 23\% & 24\%\\
\hline
$A5^{\prime}$  & 30 sec. (OFF) - 70 sec. (ON) & 4 sec. & 2\% & 10\%\\
\hline \hline

$B1$  & 50\% - 50\% & 150 sec. & 32\% & 23\%\\
\hline
$B2$  & 50\% - 50\% & 300 sec. & 26\% & 24\%\\
\hline 
\end{tabular}}
\end{table}
We considered a total of 12 cases for on-off attacks, where each simulation consists of 50 IoT devices and one on-off attacker. Table \ref{table_4} shows the simulation settings and the experimental results, depicting the reduction in trust scores achieved by the proposed scheme compared to other mechanisms in  each case. These cases vary by the service request interval of IoT devices and the ON-OFF period of the service provider. The service request interval denotes the time between successive service requests made by an IoT device. The smaller service request interval indicates that the IoT device requests service frequently which results in a large number of trust values in the time window. On the contrary, a higher service request interval results into less number of trust scores in the time window due to infrequent service requests. The ON-OFF interval specifies the times of the ON and OFF state of a service provider. For example, 70 sec. (OFF) - 30 sec. (ON) indicates that the service provider begins with the OFF state and remains in the OFF state (behaves normally as an honest service provider) for 70 seconds. During this time, 95\% of the requests served will have higher trust scores. Then it shifts to the ON state or the attack state (behaves as a malicious service provider) and remains in that state for 30 seconds. During this time frame, 95\% of the services served will have lower trust scores. After that, it again goes to the OFF state and repeats this cycle for the rest of the simulation time. The 50\%-50\% case in Table \ref{table_4} signifies that the service provider randomly serves 50\% of the total services maliciously and the rest of the 50\% of the services honestly. 

The proposed, ATMRF, and TMRF schemes behave differently in dealing with on-off attackers. As discussed in Section \ref{on-off-security}, the proposed scheme adjusts the length of the time window dynamically to ensure sufficient trust scores in the 
window even when the number of transactions is infrequent. Besides, it utilizes harmonic mean to compute $T_{intermediate}$ (Eq. \ref{Eqn_3}) to prevent drastic changes in the trust score of a service provider. It also imposes a penalty factor, $E$ for malicious behavior that increases the penalty with the number of bad transactions (Eq. \ref{Eqn_4}). Moreover, the proposed scheme remembers past behavior by combining the earlier and current domain trust scores in computing the final domain trust score. On the contrary, both ATMRF and TMRF use a fixed window length, which does not ensure sufficient trust scores in the case of rare transactions. They both emphasize recent trust scores, enabling attackers to recover quickly after malicious actions by performing just a few good transactions. Besides, both schemes utilize a static penalty factor to amplify the impact of negative feedback, overlooking the nuanced variations in the influence exerted by different types of attacks. Cases $A1 \sim B2$ in Table \ref{table_4} are designed to demonstrate the impact of the features of the proposed scheme in handling on-off attacks.
\begin{figure*}
    \centering
    
    \subfigure[$A1$: 30 sec. (ON) - 70 sec. (OFF)] {\includegraphics[width=.30\textwidth]{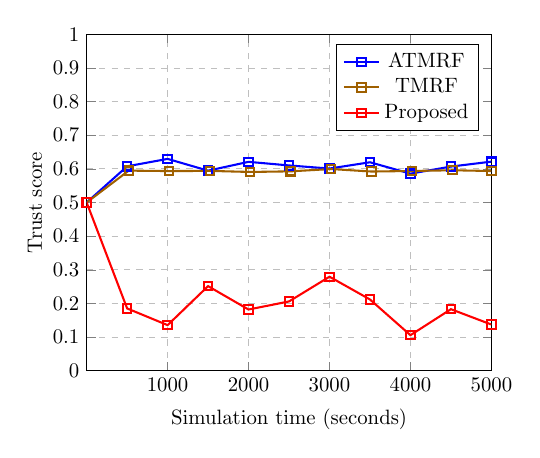}
    \label{a1}} 
    %\hfil
    \subfigure[$A1^\prime$: 70 sec. (OFF) - 30 sec. (ON)]{\includegraphics[width=.30\textwidth]{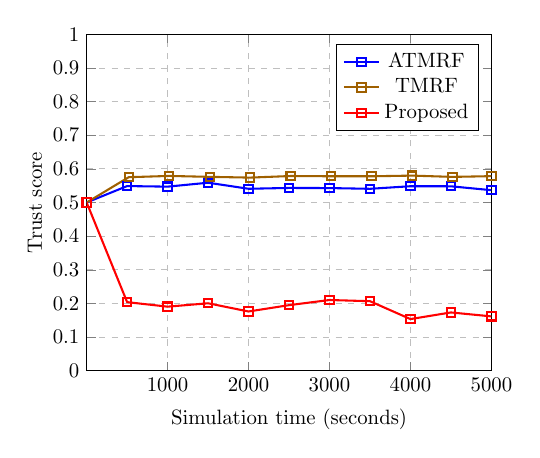}
    \label{a1p}}
    \subfigure[$A2$: 40 sec. (ON) - 60 sec. (OFF)] {\includegraphics[width=.30\textwidth]{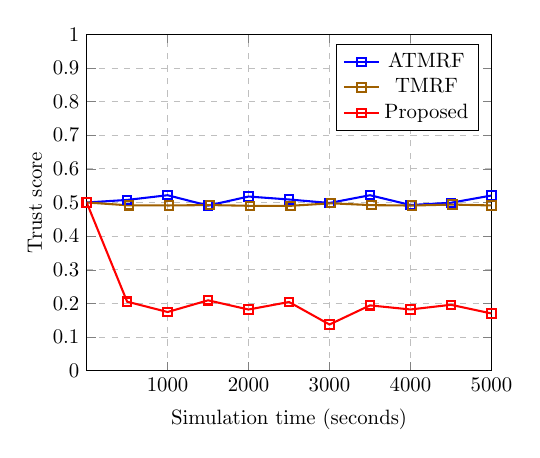}
    \label{a2}} 
    %\hfil
    \subfigure[$A2^\prime$: 60 sec. (OFF) - 40 sec. (ON)] {\includegraphics[width=.30\textwidth]{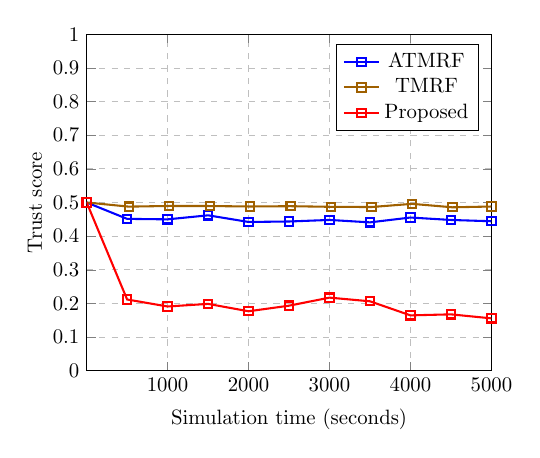}
    \label{a2p}} 
    \subfigure[$A3$: 50 sec. (ON) - 50 sec. (OFF)] {\includegraphics[width=.30\textwidth]{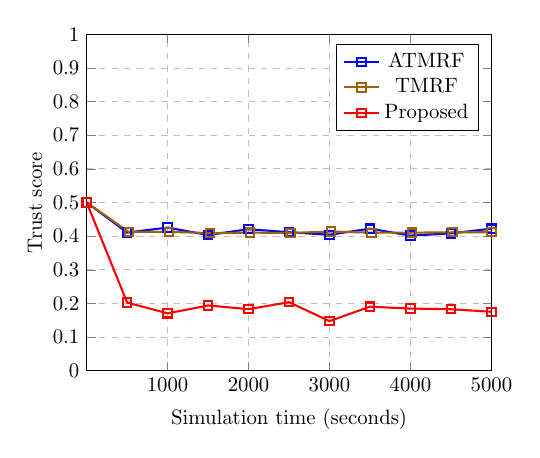}
    \label{a3}} 
    %\hfil
    \subfigure[$A3^\prime$: 50 sec. (OFF) - 50 sec. (ON)] {\includegraphics[width=.30\textwidth]{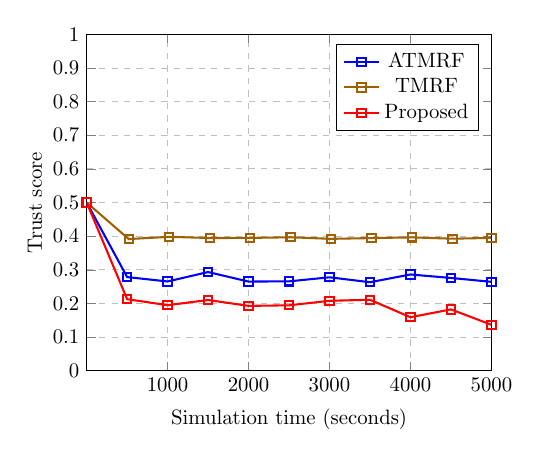}
    \label{a3p}}  
    \subfigure[$A4$: 60 sec. (ON) - 40 sec. (OFF)] {\includegraphics[width=.30\textwidth]{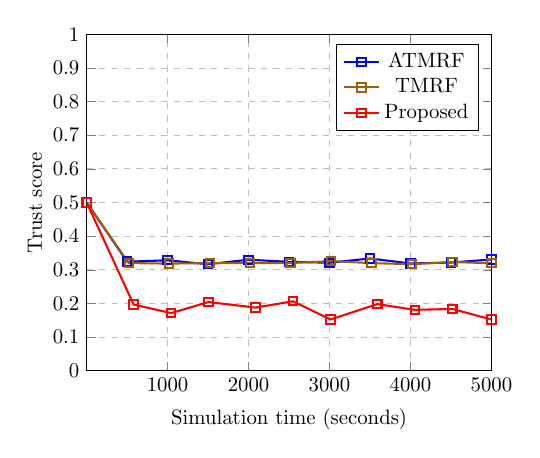}
    \label{a4}} 
    %\hfil
    \subfigure[$A4^\prime$: 40 sec. (OFF) - 60 sec. (ON)]{\includegraphics[width=.30\textwidth]{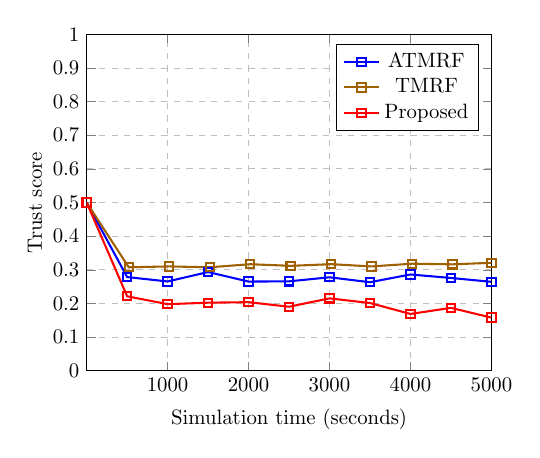}
    \label{a4p}}
    \subfigure[$A5$: 70 sec. (ON) - 30 sec. (OFF)] {\includegraphics[width=.30\textwidth]{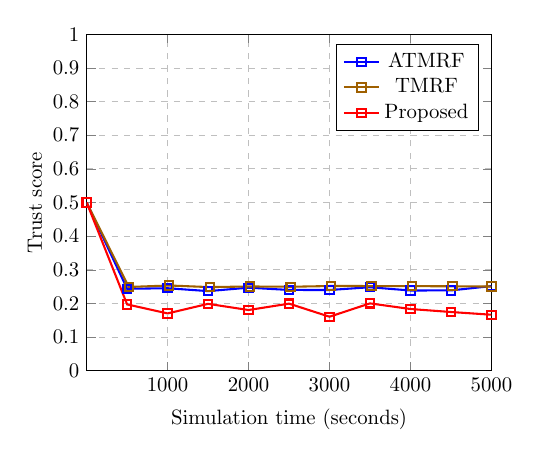}
    \label{a5}} 
    %\hfil
    \subfigure[$A5^\prime$: 30 sec. (OFF) - 70 sec. (ON)] {\includegraphics[width=.30\textwidth]{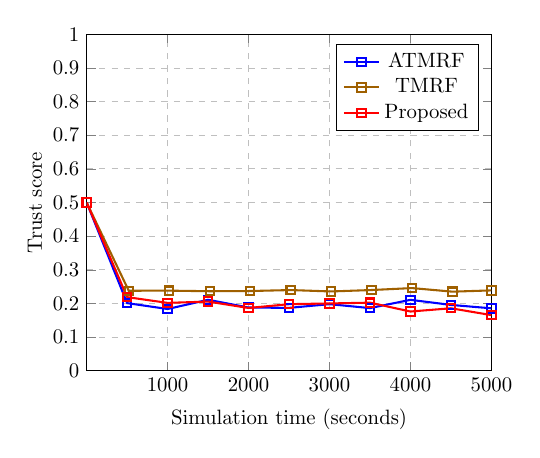}
    \label{a5p}} 
    \subfigure[$B1$: 50\% - 50\%] {\includegraphics[width=.30\textwidth]{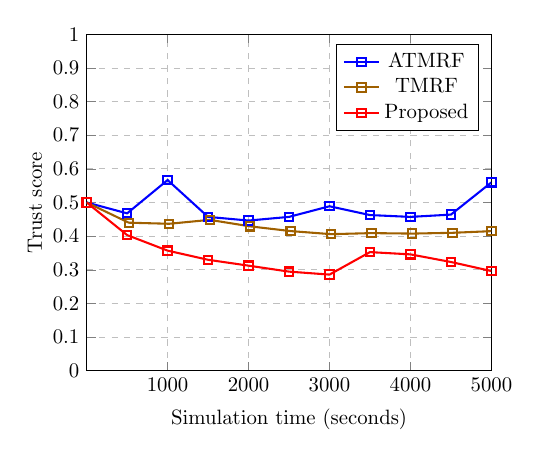}
    \label{b1}} 
    %\hfil
    \subfigure[$B2$: 50\% - 50\%] {\includegraphics[width=.30\textwidth]{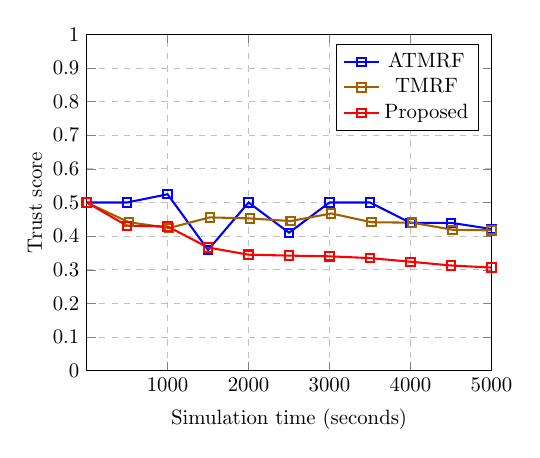}
    \label{b2}}
    \caption{Trust value convergence of an on-off attacker.}
    \vspace{-1.2em}
    \label{fig_15}
\end{figure*}

Figure \ref{fig_15} shows the results obtained from the simulations mentioned in Table \ref{table_4}. Cases $A1 \sim A5^\prime$ were designed to demonstrate the effectiveness of the direct trust calculation procedure (Eq. \ref{Eqn_3} and Eq. \ref{Eqn_4}) of the proposed scheme. In cases $A1 \sim A5^\prime$, the service request interval was 4 seconds, supporting a sufficient number of ratings in the time windows of each IoT device in every scheme. The ON and OFF states of the attacker were varied to show the impact of direct trust computation in the proposed scheme. We observe that the proposed scheme computes domain trust score (Eq. \ref{Eqn_5}) of the attacker more accurately in every case as shown in Table \ref{table_4}. Averaging the decrease rates mentioned in Table \ref{table_4} yields that the proposed scheme exhibits nearly 42\% more accuracy against ATMRF and 44\% against TMRF. Besides, for ON-OFF configurations ($A1, A2, A3, A4, A5$), the gap between the trust scores computed by the proposed scheme and the ATMRF increases compared to the respective OFF-ON configurations ($A1^{\prime}, A2^{\prime}, A3^{\prime}, A4^{\prime}, A5^{\prime}$), visible from Figs. \ref{a1}, \ref{a1p}, \ref{a2}, \ref{a2p}, \ref{a3}, \ref{a3p}, \ref{a4}, \ref{a4p}, \ref{a5}, and \ref{a5p}. ATMRF assigns more weight to recent trust scores in a time window. Hence, ATMRF increases the trust scores of a service provider by assigning more weights to the recent trust scores in the ON-OFF configurations. For example, in case $A1$: 30 sec. (ON) - 70 sec. (OFF), ATMRF assigns more weight to 70 sec. (OFF) behavior, increasing the trust score of the service provider compared to that of case $A1^{\prime}$: 70 sec. (OFF) - 30 sec. (ON). Although TMRF demonstrates a similar trend to ATMRF by assigning higher trust scores in ON-OFF configurations compared to OFF-ON configurations, the magnitude of change is comparatively lower due to its more conservative trust update mechanism. While ATMRF replaces the exact transaction time with average timestamp per time slot causing abrupt changes in the decay values, TMRF maintains smoother decay by accounting for each transaction's exact time. In contrast, the proposed scheme exhibits a consistent trust score in both OFF-ON and ON-OFF configurations due to the use of $T_{intermediate}$, dynamic penalty factor $E$, and the capability of remembering the past behavior of a service provider.

We also observed that for ON-OFF configurations ($A1$, $A2$, $A3$, $A4$, and $A5$), both ATMRF and TMRF show improved accuracy in identifying an on-off attacker with increasing ON duration followed by decreasing OFF duration. With increasing duration of the ON period in the ON-OFF configurations, the trust scores for malicious behavior during the ON period gradually receive significant weight, depicting more accuracy in identifying an attacker in ATMRF and TMRF. For example, when comparing $A1$: 30 sec. (ON) – 70 sec. (OFF) in Fig. \ref{a1} with $A5$: 70 sec. (ON) – 30 sec. (OFF) in Fig. \ref{a5}, it is evident that Fig. \ref{a5} demonstrates higher accuracy in detecting on-off attackers for ATMRF and TMRF. 

Similarly, in the OFF-ON configurations ($A1^\prime$, $A2^\prime$, $A3^\prime$, $A4^\prime$ and $A5^\prime$), we notice that the increasing ON period lower the curve of trust scores of ATMRF and TMRF as low trust scores in the ON period are getting higher weights in computing the trust score of the attacker, visible from Figs. \ref{a1p}, \ref{a2p}, \ref{a3p}, \ref{a4p}, and \ref{a5p}.

To illustrate the effectiveness of the dynamic length-adjusting window mechanism, we designed cases $B1$ and $B2$. In both cases, the service provider provided service in such a manner that 50\% of the total service was delayed, and the remaining 50\% of the service was served within the threshold. In case $B1$, where the service request interval was 150 seconds, the proposed scheme attains approximately 32\% and 23\% more accuracy than ATMRF and TMRF, respectively, visible from Fig. \ref{b1}. In case $B2$, where the service request interval was 300 seconds, the proposed scheme reduces the trust values by approximately 26\% and 24\% compared to ATMRF and TMRF, respectively, shown in Fig. \ref{b2}. 
In both $B1$ and $B2$, the IoT devices in ATMRF and TMRF have inadequate ratings stored in their windows. With little to no ratings stored in the windows, IoT devices designated the on-off attacker in the uncertain zone in ATMRF and TMRF. On the contrary, the proposed scheme utilizes the dynamic length-adjusting window mechanism, allowing IoT devices to store the minimum number of ratings, which results in a more accurate calculation of the domain trust score.   

\begin{figure}
    \centering
    \includegraphics[width=.45\textwidth]{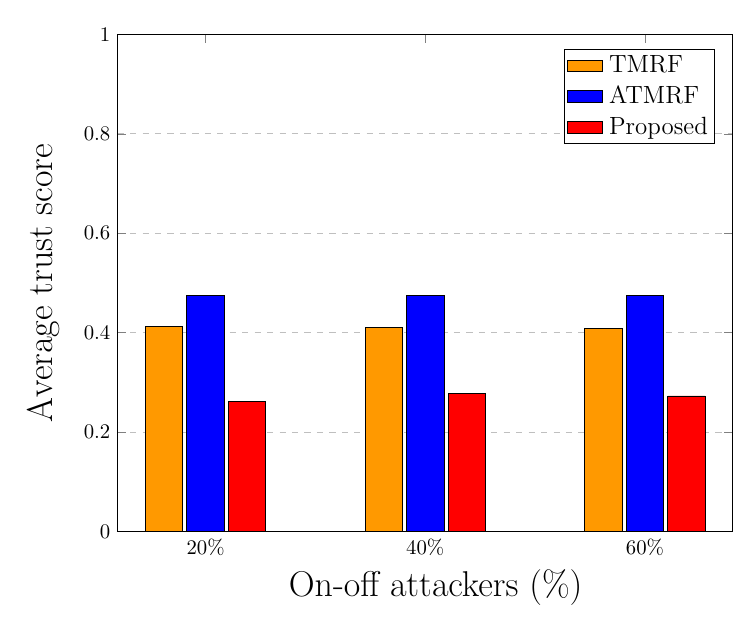}
    \caption{Performance analysis for varying percentage of on-off attackers.}
    \label{fig_16}
\end{figure}

To demonstrate the robustness and adaptability of the proposed scheme in handling on-off attacks, the performance of the proposed scheme is analyzed with an increasing percentage of attackers. For this experiment, we considered 50 IoT devices and five service providers, with a varying percentage of SP acting as on-off attackers. We considered scenarios where 20\%, 40\%, and 60\% of the five SPs were on-off attackers. The service request interval was 4 seconds and the on-off attackers randomly served 50\% of the services honestly and the other 50\% maliciously. We computed the trust score of each on-off attacker over a duration of 5000 seconds and then averaged these scores to determine the mean trust score for the on-off attackers during that time. Finally, we averaged the trust scores over $5000$ seconds to find the overall average trust score for a specific percentage of on-off attackers. Figure \ref{fig_16} shows that the proposed scheme consistently detects on-off attackers more accurately across all scenarios compared to ATMRF and TMRF.

\subsection{Performance Analysis in Mixed-Attack Scenarios}\label{sec_mixed_scenario}
The proposed scheme's performance is evaluated in mixed-attack scenarios where multiple attacks occur simultaneously.

\subsubsection{On-off, bad-mouthing, and ballot-stuffing attacks occur concurrently}\label{mixed_attack_scenario}
For this experiment, we considered 50 IoT devices and 3 service providers. Among the 50 IoT devices, forty behaved honestly, five conducted bad-mouthing attacks, and the remaining five performed ballot-stuffing attacks. Among the service providers, one worked honestly, one was malicious (providing low-quality services), and the other performed on-off attacks (50\%-50\% behavior). Each IoT device requested services every four seconds.
\begin{figure*}
    \centering
    \subfigure[Honest service provider] {\includegraphics[width=.30\textwidth]{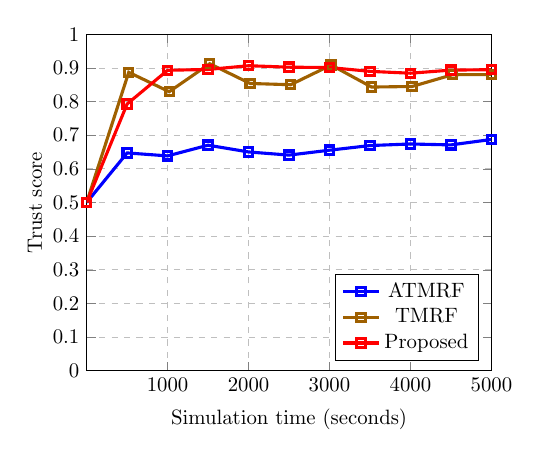}
    \label{fig_17a}} 
    %\hfil
    \subfigure[Malicious service provider]{\includegraphics[width=.30\textwidth]{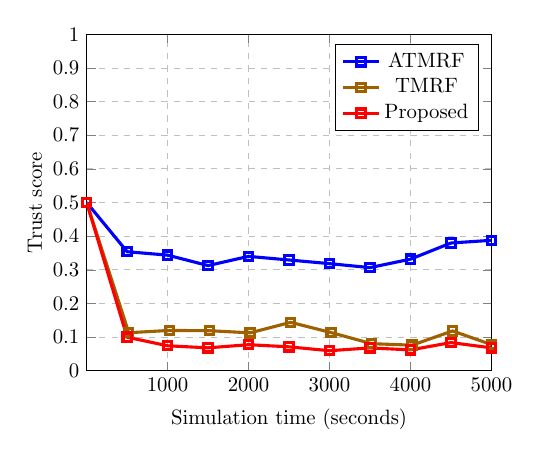}
    \label{fig_17b}}
    \subfigure[On-off service provider] {\includegraphics[width=.30\textwidth]{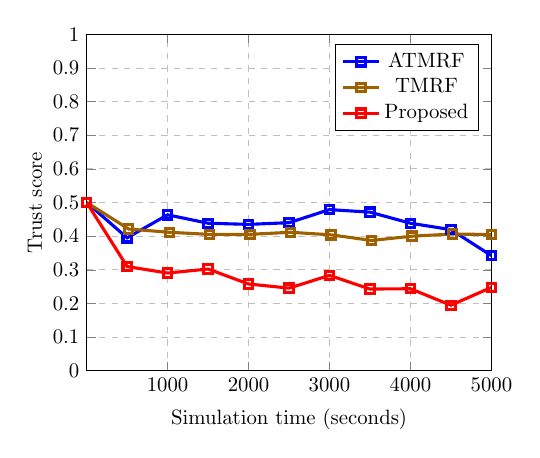}
    \label{fig_17c}} 
    \caption{Trust convergence in presence of on-off, bad-mouthing and ballot-stuffing attacks.}
    \vspace{-1.2em}
    \label{fig_17}
\end{figure*}

Figure \ref{fig_17} demonstrates that the proposed scheme identifies honest SPs, malicious SPs, and the on-off attackers more accurately than ATMRF. The proposed scheme generates trust scores of approximately 0.9 for honest SP, trust scores of less than 0.1 for malicious SP, and trust scores between 0.2 $\sim$ 0.3 for on-off attackers after 1000 seconds, as illustrated in Fig. \ref{fig_17a}, \ref{fig_17b}, \ref{fig_17c}, respectively. The proposed scheme exhibits a consistent increase in the trust score for the honest SP due to the preservation of sufficient trust scores in the time window, the use of $T_{intermediate}$, the application of equal weight to the previous and new domain trust scores, the utilization of efficient reward functions in Eq. \ref{Eqn_4}, and the ability to successfully filtering bad trust scores (bad-mouthing attacks) using subspace clustering. Besides, the proposed scheme effectively penalizes malicious actions through the penalty function in Eq. \ref{Eqn_4}, successfully filters out ballot-stuffing attacks using subspace clustering, and effectively forwards prior trust evaluation to subsequent calculations. Furthermore, the proposed scheme effectively identifies and penalizes on-off SPs. It ensures sufficient trust scores inside the time window, uses $T_{intermediate}$ to prevent drastic fluctuations in trust scores due to recent behavior, utilizes $R$ and $E$ in direct trust computation (Eq. \ref{Eqn_4}) and incorporates both past and present domain trust to effectively handle on-off attackers.

Figures \ref{fig_17a} and \ref{fig_17b} show that TMRF performs comparably to the proposed scheme in detecting honest and malicious SPs, as fuzzy C-Means clustering effectively eliminates bad recommendations and enhances detection accuracy. However, TMRF lacks robustness in identifying on-off SPs shown in Fig. \ref{fig_17c} due to its reliance on a time decay function that cannot adequately capture the fluctuating behavior of on-off attackers.

As shown in Fig. \ref{fig_17a}, ATMRF struggles to accurately identify honest SPs due to lack of effective reward mechanism, inability to effectively separate malicious feedback (bad-mouthing attacks) from legitimate feedback, and the lack of persistence in forwarding prior trust evaluations to subsequent calculations. Additionally, Fig. \ref{fig_17b} shows that ATMRF cannot properly identify malicious SP due to a slow penalty mechanism, inability to differentiate ballot stuffing attacks, and failure to propagate prior trust scores in the subsequent computations. Additionally, ATMRF cannot handle on-off SPs shown in Fig. \ref{fig_17c} due to the model's emphasis on recent trust scores, the use of a static penalty factor, and its inability to properly distinguish between alternating behaviors.

\subsubsection{On-off and bad-mouthing (on-off mode) attacks occur simultaneously}

We considered an attack scenario where IoT devices engaged in bad-mouthing attacks in an on-off manner and service providers executed on-off attacks. The simulation environment comprised 50 IoT devices and three service providers. Each IoT device made service requests every four seconds. Among the 50 IoT devices, 40 operated honestly, and the remaining ten participated in bad-mouthing attacks in an on-off manner. Specifically, the bad-mouthing attackers behaved honestly for the first 25 seconds, conducted attacks for the next 25 seconds, reverted to honest behavior in the subsequent 25 seconds, and then executed bad-mouthing attacks for the following 25 seconds. This cycle was repeated continuously every 100 seconds. Among the three service providers, one worked honestly, one acted maliciously, and the other exhibited on-off attack behavior. The on-off SP worked honestly for the first 25 seconds of each 100-second interval, while performing on-off attacks (50\%-50\%) for the remaining 75 seconds. This cycle was repeated every 100 seconds.

\begin{figure*}
    \centering
    \subfigure[Honest service provider] {\includegraphics[width=.30\textwidth]{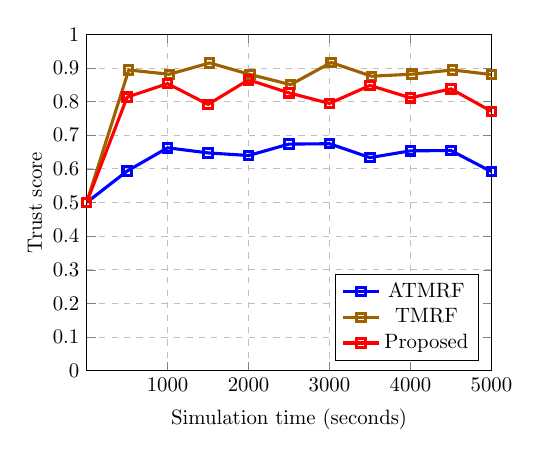}
    \label{fig_18a}} 
    \subfigure[Malicious service provider]{\includegraphics[width=.30\textwidth]{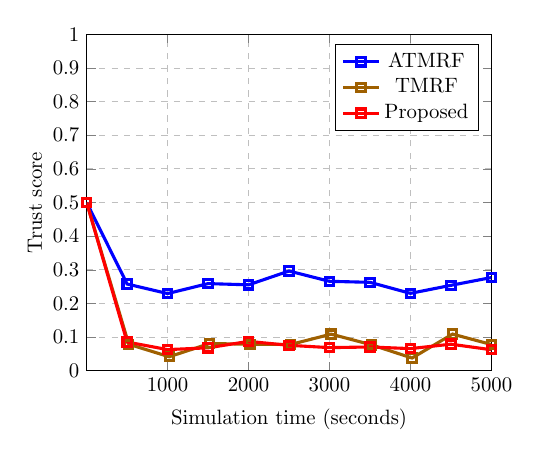}
    \label{fig_18b}}
    \subfigure[On-off service provider] {\includegraphics[width=.30\textwidth]{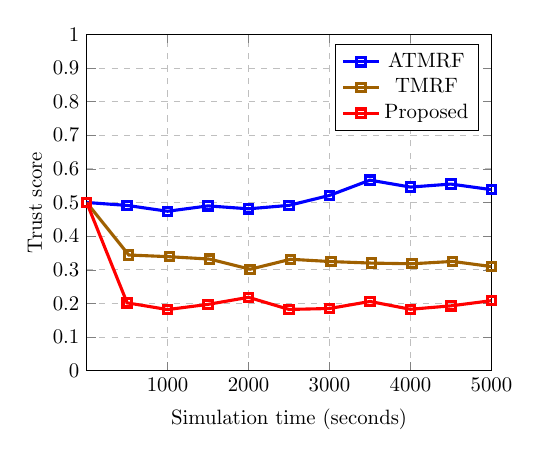}
    \label{fig_18c}} 
    \caption{Trust convergence in presence of on-off and bad-mouthing attacks (on-off mode).}
    \label{fig_18}
    
\end{figure*}

Figure \ref{fig_18} shows that the proposed scheme outperforms ATMRF in identifying the honest, malicious, and on-off SPs. The reasons behind the proposed scheme's good performance and ATMRF's poor performance are discussed in Section \ref{mixed_attack_scenario}. TMRF performs slightly better than the proposed scheme in evaluating the honest SP, shown in \ref{fig_18a} and achieves similar accuracy to the proposed scheme in identifying malicious SP shown in \ref{fig_18b},  due to its use of the Fuzzy C-Means clustering for effective recommendation filtering. Similar to ATMRF, TMRF struggles to detect the on-off attacker effectively due to its dependency on a time decay function, which does not sufficiently capture the intermittent malicious behavior of on-off SPs, shown in \ref{fig_18c}. However, Fig. \ref{fig_18} shows some discrepancies from Fig. \ref{fig_17} due to the relatively larger number of bad-mouthing attackers, their on-off mode of operations, and the absence of ballot-stuffing attacks. For instance, ATMRF shows lower trust score for malicious SP shown in Fig. \ref{fig_18b} compared to the trust score assigned to the malicious SP in Fig. \ref{fig_17b} due to the absence of ballot-stuffing attacks. In the proposed scheme, the trust score of the on-off attacker in Fig. \ref{fig_18c} is lower compared to that in Fig. \ref{fig_17c} because of the absence of ballot-stuffing attacks.

\subsection{Efficiency of cluster computation}\label{filter_efficiency}

\begin{figure}
    \centering
    \includegraphics[width=0.5\textwidth]{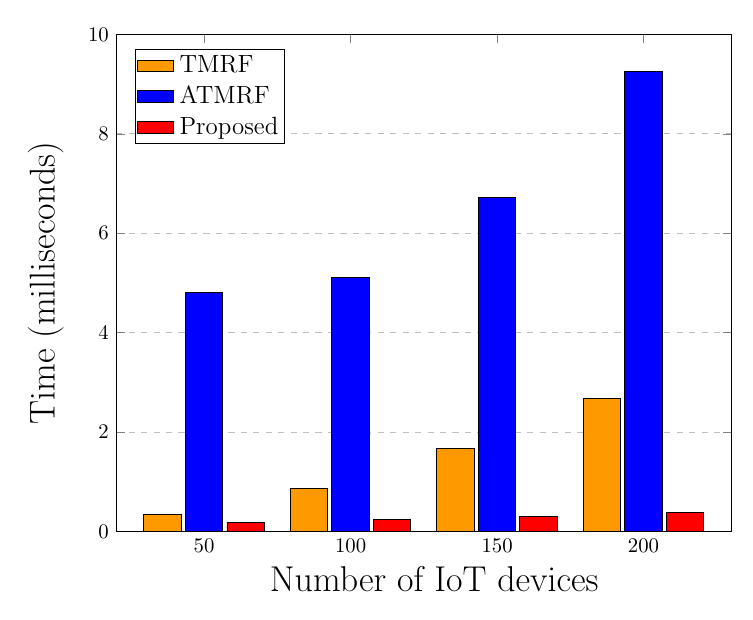}
    \caption{Time to compute clusters.}
    \label{fig_19}
\end{figure}

The proposed scheme uses a customized subspace clustering algorithm that runs in linear time complexity of $\mathcal{O}(n)$ to filter out malicious recommendations, outperforming the clustering algorithms used by ATMRF and TMRF. The ATMRF utilizes the $k$-means clustering algorithm to remove  malicious trust scores, which is computationally more expensive and has a polynomial time complexity. For large number of data points, the $k$-means algorithm is not feasible. On the other hand, the TMRF uses the Fuzzy C-Means clustering algorithm, which also has a polynomial time complexity \cite{fcm}. However, FCM outperforms $k$-means because its soft clustering prevents boundary oscillation, where data points near cluster boundaries cause $k$-means to waste iterations switching points back and forth between clusters. In contrast, the FCM allows gradual membership adjustments, leading to faster and smoother convergence. Figure \ref{fig_19} shows the time to compute clusters in all the schemes with the increasing number of IoT devices. Here note that, the number of data points to compute the clusters is equal to the number of IoT devices in each scheme. As shown in Fig. \ref{fig_19}, the proposed scheme computes clusters in $0.1 \sim 0.3$ milliseconds depending on the number of devices. On the contrary, the ATMRF and TMRF compute clusters in $5 \sim 9$ milliseconds and $0.3 \sim 2.7$ milliseconds, respectively. Consequently, the proposed scheme achieves nearly 95\% and 71\% reduction in clustering time compared to ATMRF and TMRF, respectively. 

%% file: conclusion.tex
\section{Conclusion}\label{sec_conclusion}
We have proposed an effective trust management scheme to secure an IoT environment from bad-mouthing, ballot-stuffing, and on-off attacks. The proposed scheme utilizes a dynamic sliding window mechanism and a domain trust computation mechanism to precisely identify on-off attackers. Besides, a modified subspace clustering algorithm is employed to remove recommendations from malicious IoT devices to prevent bad-mouthing and ballot-stuffing attacks. The experimental results demonstrate that the proposed scheme decreases the trust score of an on-off attacker by nearly 44\%, enabling better detection of on-off service providers. Moreover, it preserves its efficacy even when the proportion of on-off attackers increases and in situations where multiple attacks happen concurrently.  In addition, the proposed scheme speeds up the overall operation by reducing the clustering time by nearly 95\%. We believe the proposed scheme is a more practical and robust solution for the IoT environment, featuring improved security against attacks and greater computational efficiency. In the future, we aim to extend our work for inter-domain settings. Besides, expanding the proposed clustering algorithm to support data space divided into more than three grids is also an interesting research direction.